\pdfoutput=1 
\documentclass[a4paper,11pt]{article}

\usepackage{jinstpub} 

\usepackage{natbib} 

\title{\boldmath The Multi-Blade Boron-10-based Neutron Detector for high intensity Neutron Reflectometry at ESS}

\author[a,1]{F. Piscitelli,\note{Corresponding author.}}
\author[b,a]{F. Messi,}
\author[a]{M. Anastasopoulos,}
\author[a]{T. Bry\'s,}
\author[a]{F. Chicken,}
\author[c,a]{E. Dian,}
\author[d]{J. Fuzi,}
\author[a,e]{C. H\"{o}glund,}
\author[d]{G. Kiss,}
\author[d]{J. Orban,}
\author[d]{P. Pazmandi,}
\author[a,e]{L. Robinson,}
\author[d]{L. Rosta,}
\author[a,e]{S. Schmidt,}
\author[d]{D. Varga,}
\author[d]{T. Zsiros,}
\author[a,f]{R. Hall-Wilton}

\affiliation[a]{European Spallation Source ERIC (ESS), \\P.O. Box 176, SE-22100 Lund, Sweden.}
\affiliation[b]{Nuclear Physics Department, Lund University, \\P.O. Box 118, SE-22100 Lund, Sweden.}
\affiliation[c]{Hungarian Academy of Sciences Centre for Energy Research, \\Konkoly Thege Mikl\'os \'ut 29-33, H-1121 Budapest, Hungary.}
\affiliation[d]{Wigner Research Centre for Physics, \\Konkoly Thege Mikl\'os \'ut 29-33, H-1121 Budapest, Hungary.}
\affiliation[e]{Thin Film Physics Division, \\Link\"{o}ping University, SE-58183 Link\"{o}ping, Sweden.}
\affiliation[f]{Mid-Sweden University, \\SE-85170 Sundsvall, Sweden.}

\emailAdd{francesco.piscitelli@esss.se}

\abstract{The Multi-Blade is a Boron-10-based gaseous detector introduced to face the challenge arising in neutron reflectometry at pulsed neutron sources. Neutron reflectometers are the most challenging instruments in terms of instantaneous counting rate and spatial resolution. This detector has been designed to cope with the requirements set for the reflectometers at the upcoming European Spallation Source (ESS) in Sweden. Based on previous results obtained at the Institut Laue-Langevin (ILL) in France, an improved demonstrator has been built at ESS and tested at the Budapest Neutron Centre (BNC) in Hungary and at the Source Testing Facility (STF) at the Lund University in Sweden. A detailed description of the detector and the results of the tests are discussed in this manuscript.}


\begin{document}
\maketitle
\flushbottom

\section{Introduction}
In this manuscript we report about an improved design of the Multi-Blade detector for neutron relfectometry applications. The general performance of the Multi-Blade detector, particularly the spatial resolution and the counting rate capability, as described in~\cite{MIO_MB2014,MIO_MyThesis} and summarized in the Table~\ref{tabref}, are confirming that this is the right way to go for a future detector technology.
\\ The Multi-Blade design has been improved and a demonstrator has been built at the European Spallation Source (ESS~\cite{ESS}). It has been tested at the Source Testing Facility (STF~\cite{SF1}) at the Lund University in Sweden and on a beam line at the Budapest Neutron Centre (BNC~\footnote{BNC is a consortium of the Centre for Energy Research operating the 10 MW research reactor and the Wigner Reserach Centre for Physics operating there the neutron scattering facilities}~\cite{FAC_BNC,FAC_BNC2,FAC_BNC3}) in Hungary. A detailed description of the detector and the results of the tests are discussed in this manuscript. 
\subsection{Detector requirements for Reflectometry instruments}
The Multi-Blade~\cite{MIO_MB2014,MIO_MyThesis} detector has been introduced to face the challanges arising in neutron reflectometry~\cite{R_NR,R_fermi}. The European Spallation Source (ESS)~\cite{ESS} is designed to be the world's brightest neutron source. The expected instantaneous neutron flux on detectors at ESS will be without precedent~\cite{ESS_TDR,DET_rates,HE3S_kirstein} and neutron reflectometers are the most challenging instruments. ESS will have two of these, FREIA~\cite{INSTR_FREIA} (horizontal reflectometer) and ESTIA~\cite{INSTR_ESTIA,INSTR_ESTIA1}  (vertical reflectometer), and the expected instantaneous local flux at the detector is between $10^{5}$ and $5\cdot10^{5}/s/mm^2$. The problem with count rates in reflectometry is a general one~\cite{HE3S_cooper}, and the ESS solution could potentially be applied to existing instruments at other neutron sources. 
\\ Along with the rate capability, the spatial resolution ($\approx 0.5mm$ needed in one direction only) is another crucial aspect that needs improvements in reflectometer detectors. There is a great interest in expanding the neutron reflectometry technique beyond static structural measurements of layered structures to kinetic studies~\cite{R_Cubitt1}. At current facilities (pulsed and reactor sources) the time resolution for kinetic studies is limited by the available flux. In references~\cite{R_Cubitt1,R_Cubitt2} a new instrument layout is presented for reactor sources to open the possibility of sub-second kinetic studies, however this requires very high spatial resolution detectors. 
\\These needs can in general not be met with the currently available technologies. $\mathrm{^3He}$ and $\mathrm{^6Li}$-scintillators are the two main technologies used in neutron detectors for reflectometry instruments. Due to the limited counting rate capability, scintillators are in general a secondary choice for high flux reflectometers and they are used as support detectors. In terms of counting rate capability, neutron detection efficiency and $\gamma$-ray sensivity~\cite{MG_khap}, $\mathrm{^3He}$ is superior to scintillators. Therefore, most existing reflectometers~\cite{INSTR_bioref,INSTR_FIGARO,INSTR_D17,INSTR_ISIS_R,INSTR_jparc,INSTR_OPAL,INSTR_REFSANS,INSTR_superadam} use this technology. Gaseous detector have several designs and performances. For cold neutrons (2.5-30\AA), $\mathrm{^3He}$ detectors have efficiencies of $50-90\%$~\cite{DET_ravazzani,HE3_thesisEd}, global count rates between $20\,KHz$ and $30\,MHz$ over the whole detector active area ($<0.5m^2$). $\mathrm{^3He}$ detectors are mainly proportional counters or Multi Wire Proportional Chambers (MWPC). Local count rates in MWPC can in principle go up to $10KHz/mm^2$~\cite{MPGD_prospect} but it has never officially been reported in the literature for neutrons. The spatial resolution that can achieved with the $\mathrm{^3He}$ technology is about $1.5\,mm$. Although the quantity of $\mathrm{^3He}$ needed for reflectometers at ESS would be available~\cite{HE3S_kouzes,HE3S_shea}, the requirements in spatial resolution and counting rate capability described above can not be fulfilled with this technology. The rate capability and spatial resolution requirements at ESS exceed the performance of current $\mathrm{^3He}$ technology by a factor of $10-100$ and a factor of $3$ respectively. On the other hand, the spatial resolution that can be achieved with Wavelength Shifting Fiber (WLS)~\cite{SCI_katagari,SCI_nakamura} detectors is below $1\,mm$ and can easily fulfil the ESS requirements for reflectometers. However, scintillators can at best reach the same counting rate capability as $\mathrm{^3He}$ and they are therefore not considered a good alternative as the main detector technology for reflectometers at ESS. 
\\ Table~\ref{tab1} summarizes the main requirements for the two neutron reflectometers at ESS. 
\begin{table}[htbp]
\centering
\caption{\label{tab1} \footnotesize Detector requirements for neutron reflectometers planned for ESS.}
\smallskip
\begin{tabular}{|l|l|l|}
\hline
\hline
  & FREIA & ESTIA  \\        
\hline
\hline
wavelength range (\AA) &  2.5 - 12  & 4 - 10 \\
\hline
efficiency  &  >40\% at 2.5\AA  & >45\% at 4\AA \\
\hline
sample-detector distance (m) & 3 & 4 \\
\hline
max instantaneous rate & & \\
on detector ($KHz/mm^2$) & $100-200$  & $100-500$ \\
\hline
size x ($mm$)  & 300 & 500 \\
size y ($mm$)  & 300 & 250 \\
\hline
spatial resolution x ($mm$)  & 2.5 & 0.5 \\
spatial resolution y ($mm$)  & 0.5 & 4 \\
\hline
uniformity ($\%$)  & 5 & 5 \\
\hline
desired max window scattering  & $10^{-4}$ & $10^{-4}$ \\
 \hline
$\gamma$-sensitivity & $<10^{-6}$ & $<10^{-6}$ \\
 \hline
 \hline
\end{tabular}
\end{table}
The detectors needed are modest in size ($\leq 500\cdot300\,mm^2$), with high counting rate capability and high spatial resolution required in one direction only. Background (background neutrons and $\gamma$-rays) suppression is also an important feature that the detector should compromise with respect to the efficiency. A suitable detector shielding should prevent the detector from counting background neutrons from the environment. Depending on the $\gamma$-ray background on the instrument, a detector should provide a suitable $\gamma$-ray rejection, typically set to about $10^{-6}$~\cite{MG_gamma,MG_khap,MPGD_CrociGamma}. 
\\ One current issue of $\mathrm{^3He}$-based detectors is the thick entrance Al-window that is needed for the high pressure of the vessel. Typically $10^{-2}$ of the incoming neutrons are scattered by this window and detected in the detector~\cite{INSTR_ESTIA}. For FREIA and ESTIA the desired detector window scattering is $10^{-4}$. 
\subsection{The Multi-Blade concept}
\begin{figure}[htbp]
\centering
\includegraphics[width=.9\textwidth,keepaspectratio]{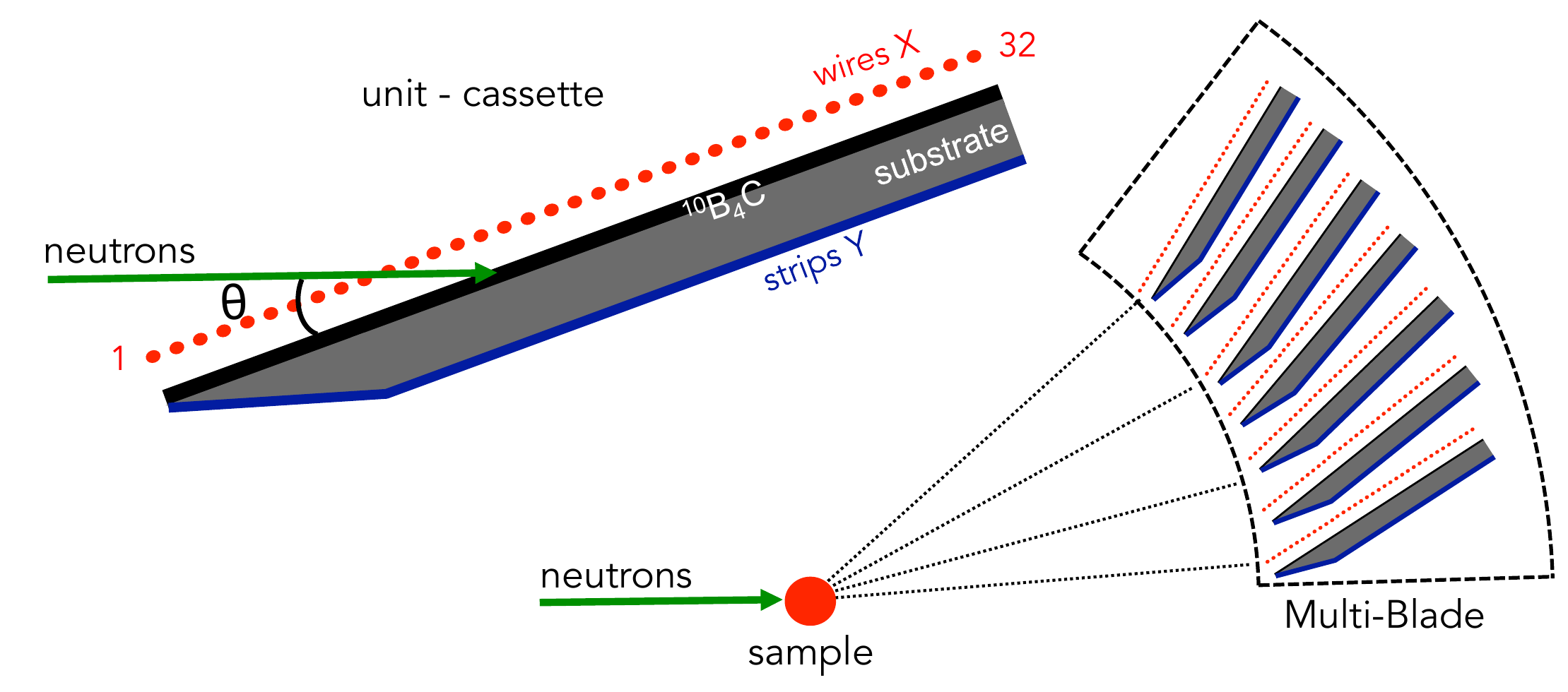}
\caption{\label{fig1} \footnotesize Schematic view of the cross-section of the Multi-Blade detector made up of identical units (cassettes) arranged one after the other. Each cassette holds a $\mathrm{^{10}B_4C}$-layer; the readout is performed through a plane of wires and a plane of strips.}
\end{figure}

The Multi-Blade concept was introduced in 2005~\cite{INSTR_Buffet2005} and two prototypes have been built in 2013 showing promising results~\cite{MIO_MB2014,MIO_MBproc,MIO_MyThesis}. We refer to this detector design as Multi-Blade 2013 and to the improved design detector presented in this manuscript as Multi-Blade 2016. The aim is to realise detectors optimized for these high rates of neutrons at ESS. A detailed description of the Multi-Blade concept can be found in~\cite{MIO_MB2014} and a sketch is shown in Figure~\ref{fig1}. 
\\ Other detector designs have been based on the same concept, an example can be found in~\cite{DET_kampmannA1CLD,DET_kampmannA1CLDp} for detectors optimized for neutron diffraction, whereas the Multi-Blade has been optimized for neutron reflectometry. 
The Multi-Blade is a stack of Multi Wire Proportional Chambers (MWPC) operated at atmospheric pressure with a continuous gas flow ($\mathrm{Ar/CO_2}$ 80/20 mixture). The Multi-Blade is made up of identical units, the so-called `cassettes'. Each cassette holds a `blade' (a flat substrate coated with $\mathrm{^{10}B_4C}$~\cite{B4C_carina,B4C_carina3,B4C_Schmidt}) and a two-dimensional readout system, which consists of a plane of wires and a plane of strips. Each $\mathrm{^{10}B_4C}$-converter (blade) is inclined at grazing angle ($\theta = 5$ degrees) with respect to the incoming neutron beam. The inclined geometry has two advantages: the neutron flux is shared among more wires with respect to the normal incidence (the counting rate capability is increased) and the spatial resolution is also improved. Moreover, the use of the $\mathrm{^{10}B_4C}$ conversion layer at an angle also increases the detection efficiency which is otherwise limited to a few percent at thermal energies for a single converter~\cite{MIO_analyt}. 
\\The proof of concept of the Multi-Blade design has been demonstrated with two prototypes~\cite{MIO_MB2014,MIO_MBproc,MIO_MyThesis} (Multi-Blade 2013). It has been shown that this detector can be operated a relatively small gas gain ($58$) and at a voltage of about $1000\,V$. Moreover, the spatial resolution has been measured and it is about $0.3\,mm\,\times4\,mm$. It has been shown that the neutron detection efficiency increases as the inclination decreases, but it has been measured at 10 degrees, while the value at 5 degrees it has been extrapolated. 
\\ A significant concern in a modular design is the uniformity of the detector response. It has been shown that an uniformity of about $2\%$ can be obtained and that the overlap between cassettes produces about a $2\,mm$ gap where the efficiency is reduced to $50\%$ with respect to the nominal efficiency. In the present prototype we are addressing these issues by reducing the gap between the cassettes. 
\\ The scattering of the materials, especially the kapton (polyimide) foil used to hold the cathode strips, used in the Multi-Blade prototype has been measured~\cite{MIO_MyThesis} and it resulted in a key point to be optimized to fulfil the ESS requirements. 
\section{Improved design of the Multi-Blade detector}\label{sect2}
\begin{figure}[htbp]
\centering
\includegraphics[width=.9\textwidth,keepaspectratio]{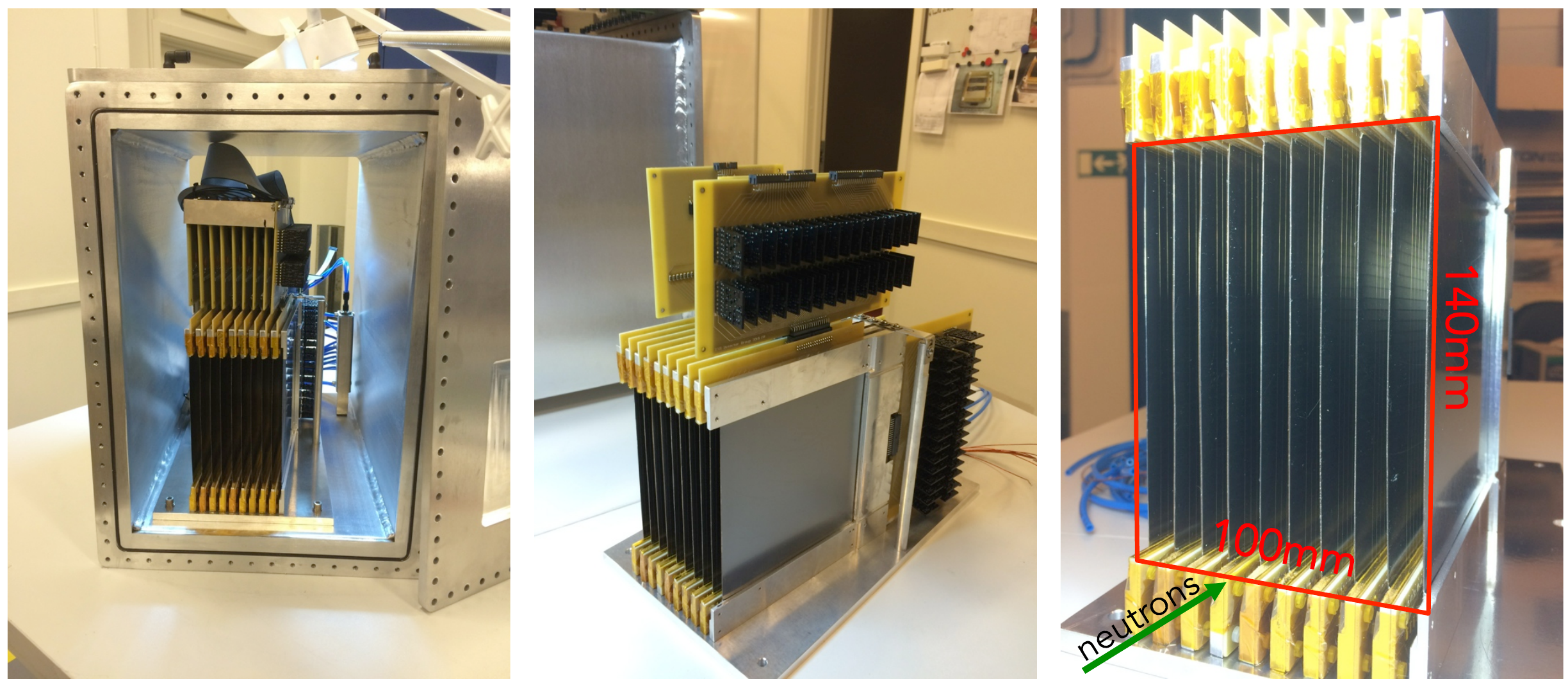}
\caption{\label{fig2} \footnotesize Three pictures of the Multi-Blade detector (Multi-Blade 2016) made of 9 cassettes.}
\end{figure}

\begin{figure}[htbp]
\centering
\includegraphics[width=.9\textwidth,keepaspectratio]{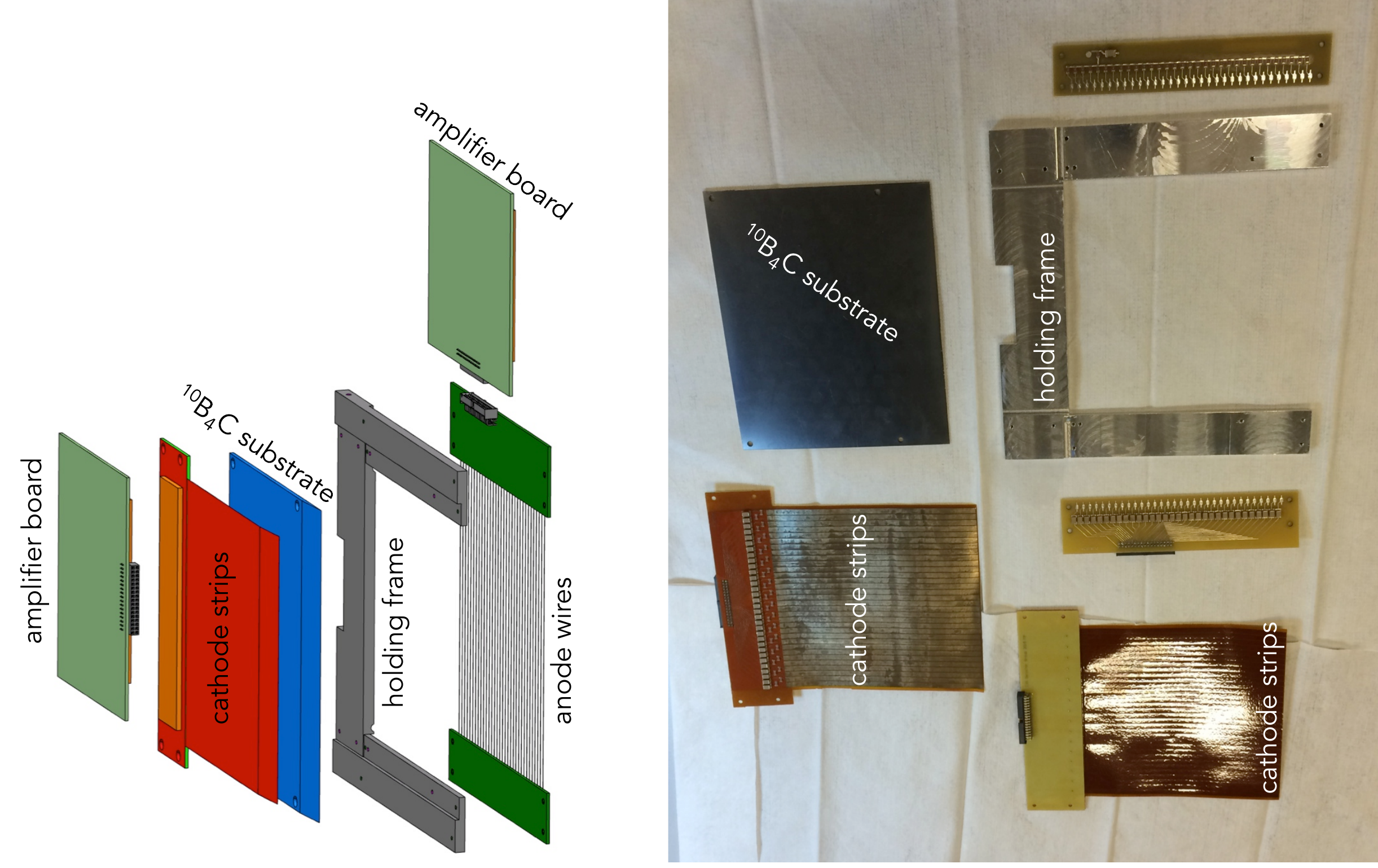}
\caption{\label{fig3} \footnotesize A CAD model of the components of a single cassette (left). The corresponding mechanical and electrical parts that make a cassette (right).}
\end{figure}

The improved version of the Multi-Blade (Multi-Blade 2016) has an active area of about $100\times 140\,mm^2$ and it is made up of nine cassettes, each one is equipped with 32 wires ($15\,\mu m$ diameter) and 32 $4mm$-wide strips (see Figure~\ref{fig2} and~\ref{fig3}). All cassettes are mechanically and electrically identical.
\subsection{Inclined geometry and single layer}
\begin{figure}[htbp]
\centering
\includegraphics[width=.7\textwidth,keepaspectratio]{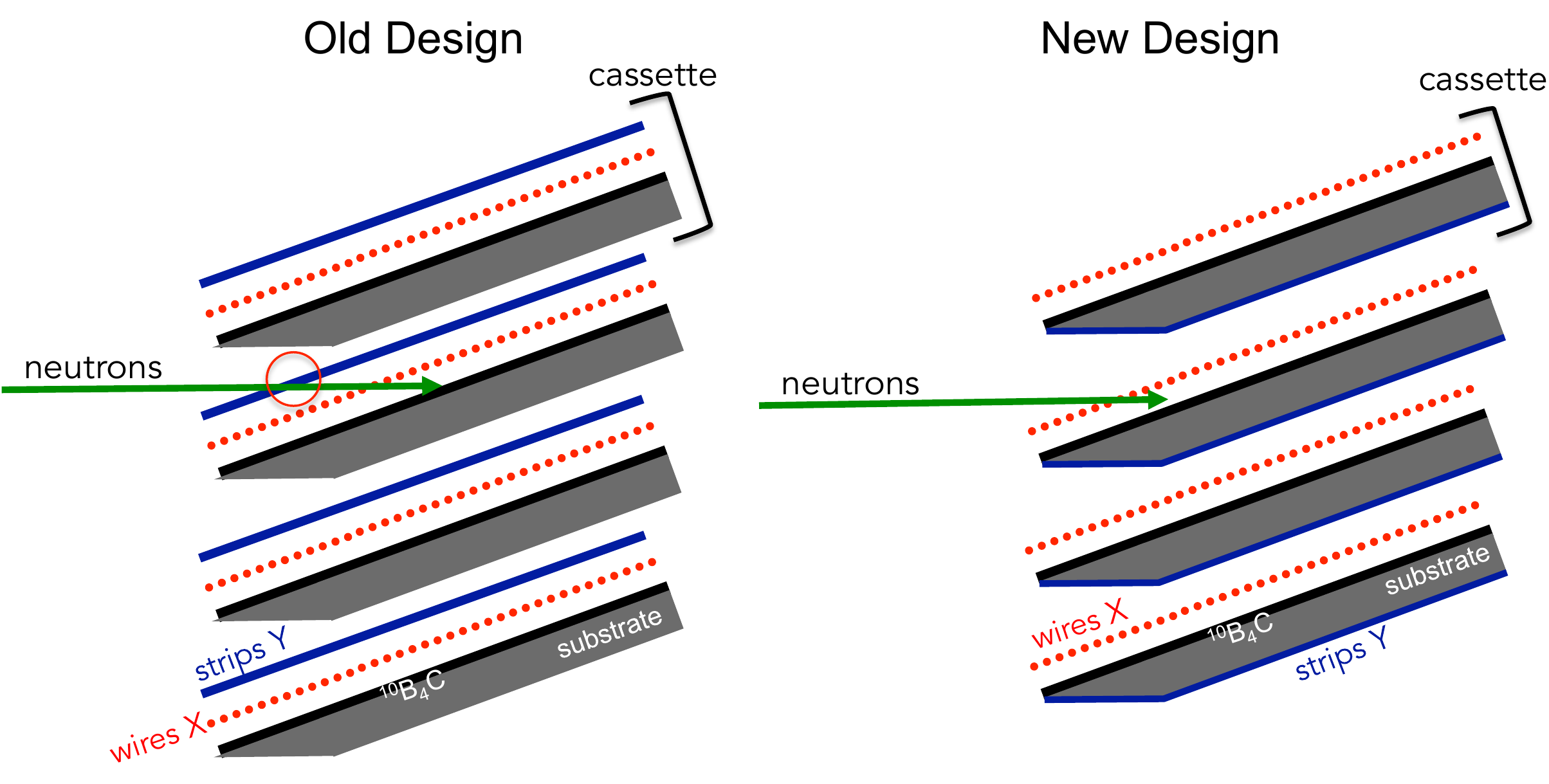}
\caption{\label{fig4} \footnotesize Old design in which the neutron have to cross the kapton foil in order to reach the converter (left). The new design where the MWPC of a cassette is completely opened to the incoming neutrons (right).}
\end{figure}

\begin{figure}[htbp]
\centering
\includegraphics[width=.49\textwidth,keepaspectratio]{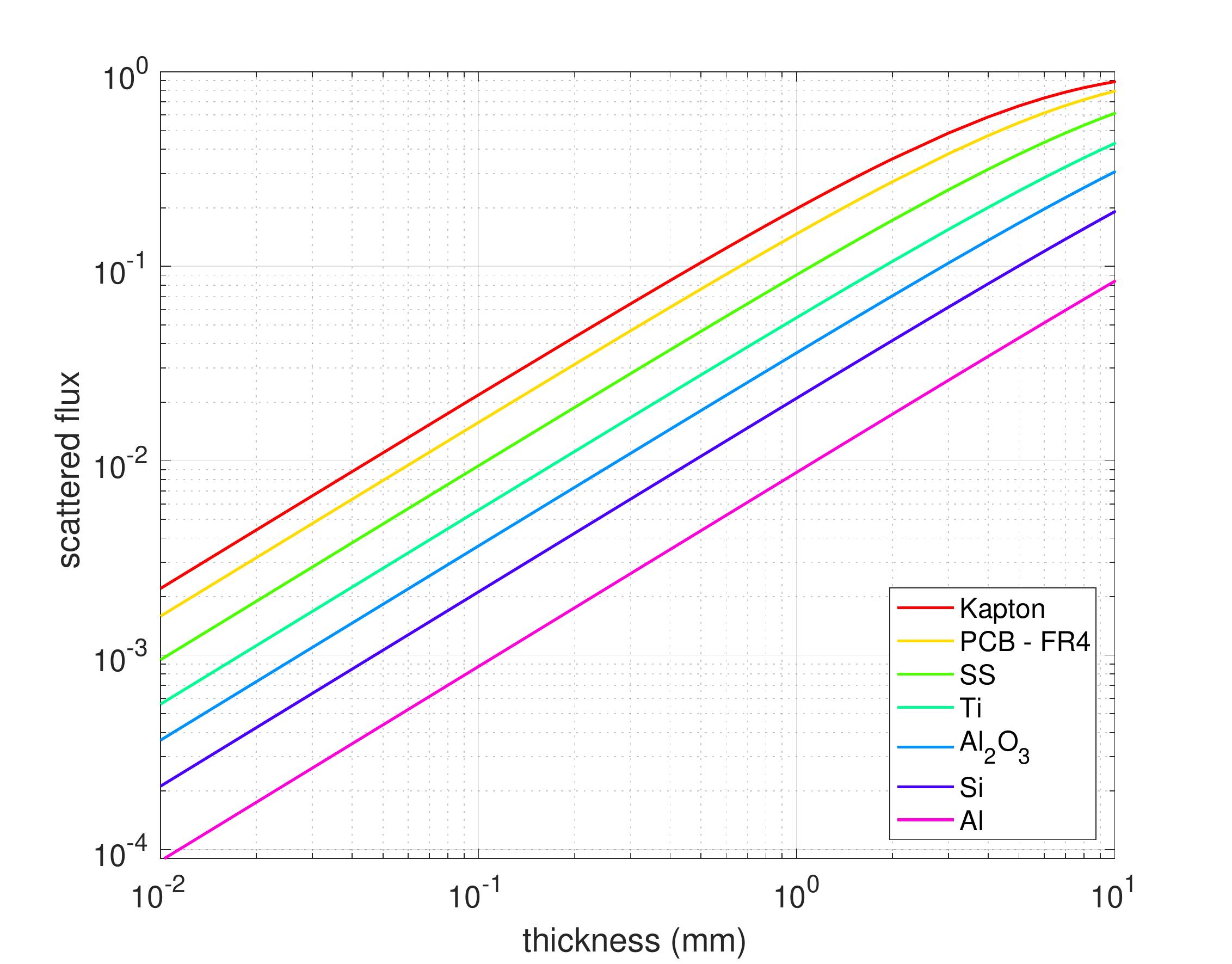}
\includegraphics[width=.49\textwidth,keepaspectratio]{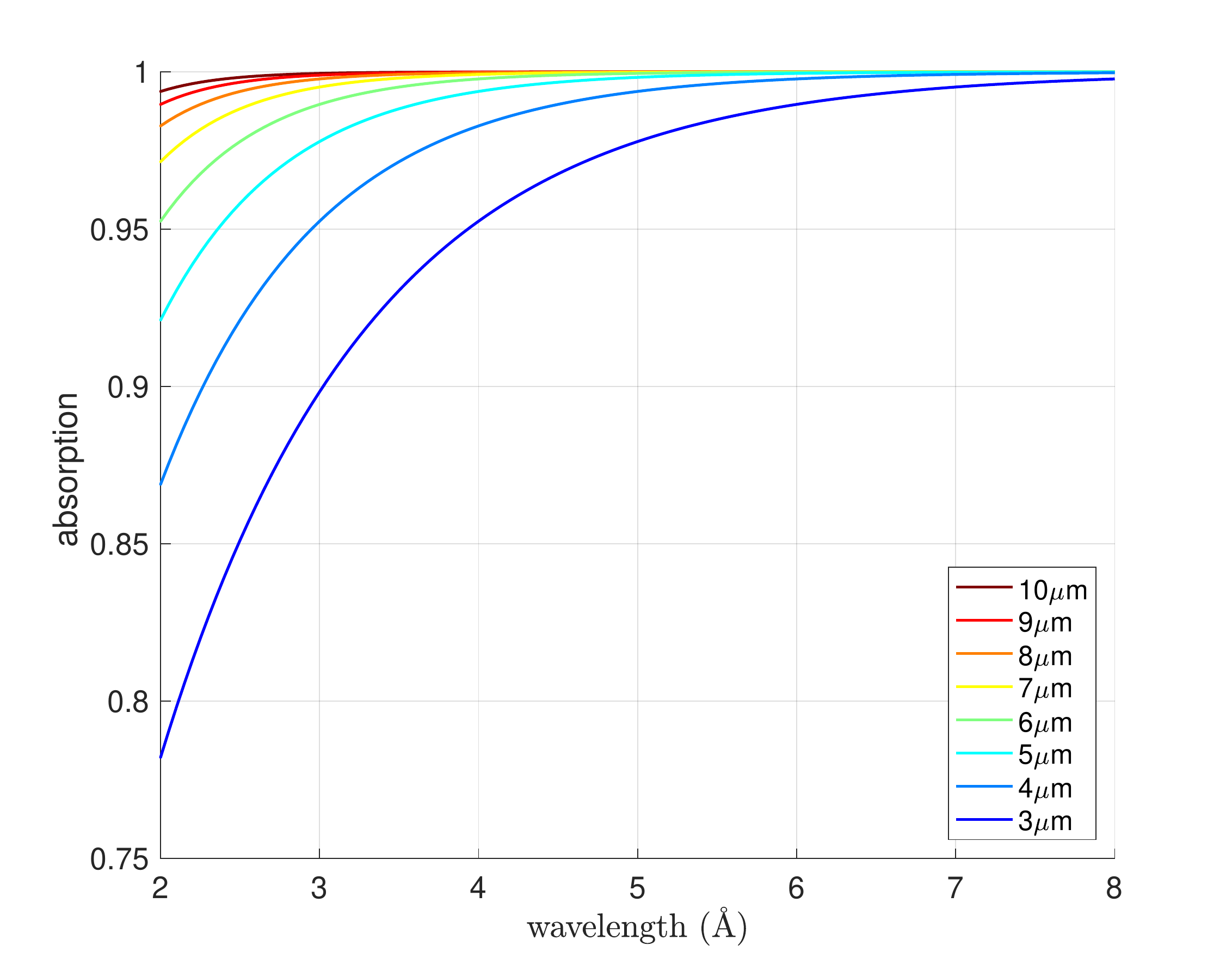}
\caption{\label{fig5} \footnotesize The fractional amount of scattered neutron flux (monochromatic at 1.8\AA) as a function of the material thickness for normal incidence (left). The fractional amount of absorbed flux by the $\mathrm{^{10}B_4C}$ coating crossed at 5 degrees for several film thicknesses as a function of the neutron wavelength (right).}
\end{figure}

The first prototype~\cite{MIO_MB2014} (Multi-Blade 2013) comprised two designs: either with one or with two $\mathrm{^{10}B_4C}$ converters: blades coated on one side or on both sides. The latter has more technical issues that makes its realization more difficult. The single layer detector is finally the choice in order to keep the mechanics reasonably simple. The 2-layer version has been rejected; the loss in the efficiency from the reduction in number of layers can be compensated by a smaller inclination but with the extra advantage of improved resolution and higher counting rate capability. 
\\ Moreover, in the two layer configuration the choice of the substrate is also crucial, due to the scattering of crossing neutrons. On the other hand, the advantage of having only one converter is that the $\mathrm{^{10}B_4C}$ coating can be of any thickness above $3\,\mu m$ without affecting the efficiency, while for the two layer option its thickness should be chosen carefully and the thickness uniformity must be controlled. Since the desired requirement for scattering is set to $10^{-4}$ the single layer option will help to solve two issues at once: the critical choice of the substrate of the neutron converter ($\mathrm{^{10}B_4C}$) and the choice of the material on which the strips lay (strip-holder). In the previous design that was a kapton (polyimide) foil that must be crossed by the neutrons in order to reach the $\mathrm{^{10}B_4C}$ layer (Figure~\ref{fig4}). Some commercial materials have been evaluated to substitute the $\mathrm{^{10}B_4C}$-substrate and the strip-holder foil in the Multi-Blade design in order to decrease the scattering; their composition is listed in Table~\ref{tab2}. 
\begin{table}[htbp]
\centering
\caption{\label{tab2} \footnotesize Composition of the materials evaluated for substrate and strip-holder in the Multi-Blade.}
\smallskip
\begin{tabular}{|l|c|l|}
\hline
\hline
Material & Density ($g/cm^3$) & Composition (fraction by weight)\\
\hline
\hline
Kapton~\cite{MATERIAL_NIST} & 1.42 & H (2.6\%), C (69\%), N (7.4\%), O (21\%) \\
(polyimide) &  &  \\
\hline
PCB - FR4~\cite{MATERIALS_FR4_1}  & 1.9 & epoxy (44\%), glass fiber (56\%) \\
 & & or H (1.4\%), C (17\%), O (43\%),  \\
 & & Mg (11.9\%), Al (13.1\%), Si (13.6\%) \\
\hline
Stainless Steel~\cite{MATERIALS_steel} & 8.03 & Fe (72\%), Ni (10\%), Cr (18\%)\\
\hline
Titanium  & 4.5 & Ti (100\%)\\
\hline
$\mathrm{Al_2O_3}$~\cite{MATERIAL_NIST} & 3.97 & O (47\%), Al (53\%)\\
\hline
Silicon   & 2.33 & Si (100\%)\\
\hline
Aluminium & 2.7 & Al (100\%)\\
\hline
\hline
\end{tabular}
\end{table}
\\ Figure~\ref{fig5} (left) shows the fractional amount of a unity neutron beam (monochromatic at 1.8\AA) scattered by a layer made of the materials listed in Table~\ref{tab2} as a function of their thicknesses. Only the scattering cross-sections~\cite{MISC_ILLblue} (coherent and incoherent) is used in this calculation since any neutron that is absorbed cannot cause any spurious event in the detector. We must consider now that these substrates and strip-holder are inclined at 5 degrees, then the actual thickness of any of these materials crossed by neutrons is a factor 11.5 larger ($1/\sin{\left(5^{\circ}\right)}=11.5$). The kapton thickness in the previous design (Multi-Blade 2013) was about $25\,\mu m$ corresponding to about $300\,\mu m$ at 5 degrees; this implies approximately $7\%$ of scattering. This value was also confirmed by the measurements carried out in~\cite{MIO_MyThesis}. From Figure~\ref{fig5} we can conclude that, because of the inclination, any chosen material for the strip-holder will cause an amount of scattering which is much larger than the requirement that has been set. 
\\ Moreover, the substrate where the $\mathrm{^{10}B_4C}$ is deposited must have some mechanical strength to sustain the residual stress of the coating~\cite{MIO_MB2014}. In the previous design a $2\,mm$-thick Al-substrate was used. For coatings deposited on a single side of the substrate, a deformation of the substrates was observed. Note that the planarity of the substrate is crucial for the uniformity of the electric field of the MWPC, and thus for the uniformity of the detector response. A simulation of the electric field have been performed in order to understand the mechanical deviations that can be accepted; they will be discussed in Section~\ref{simu}. In the current detector (Multi-Blade 2016) a $7.5\mu m$-thick $\mathrm{^{10}B_4C}$-layer has been deposited on one side of a $2\,mm$-thick Aluminium substrate. 
Figure~\ref{fig5} (right) shows the amount of absorbed neutron flux by the $\mathrm{^{10}B_4C}$-layer inclined at 5 degrees as a function of the neutron wavelength for several layer thicknesses. Note that the efficiency is saturated above $3\,\mu m$ and any extra film thickness will only help to absorb neutrons; e.g. a film of $5\,\mu m$ will absorb more than $95\%$ of the neutrons at the shortest wavelength interesting for the ESS neutron reflectometers (2.5\AA). This means that only about $5\%$ of the neutrons (at the shortest wavelength) can reach the substrate and thus can be scattered. If the probability of scattering from the substrate is e.g. $10\%$ (see Figure~\ref{fig5}) the maximum possible scattering is only $0.5\%$ at the shortest wavelength.
\\Referring to Figure~\ref{fig4}, in the new design we decided to keep the kapton as the material for the strip-holder for the simplicity of its realization. Both the substrate and the strip-holder are hidden behind a $7.5\mu m$ thick $\mathrm{^{10}B_4C}$-layer; the converter absorption in this case is $99\%$ at 2.5\AA. In this configuration the substrate and strip-holder material is not crucial for scattering since is almost not crossed by neutrons. This is finally the choice to lower the scattering within the detector at the price of a more complicated electric field geometry in the MWPC. The readout of a single $\mathrm{^{10}B_4C}$-converter is then performed by the facing anode wire plane and the strip plane that mechanically belongs to the adjacent cassette instead. 
\subsection{Multi Wire Proportional Chamber choice}
\subsubsection{Multi Wire Proportional Chamber rate capability considerations}
Due to good position resolution, low material budget and low cost MWPCs are widely used in high-energy physics~\cite{RC_andronic}. It is well known that in proportional counters and MWPC the voltage output depends on the count rate and the effect is explained by space charge (the presence of slowly moving positive ions) which decreases the electric field near the anodes thus decreasing the gas amplification factor~\cite{RC_sipila}. The MWPC geometry in the Multi-Blade has been optimized in order to reach the needed counting rate capability and spatial resolution. The three main aspects of this MWPC are: the small gap between wires and strips, the low gas gain operation and the inclined geometry. 
\\ The spatial resolution is limited by the track lengths of the neutron capture reaction fragments in $\mathrm{Ar/CO_2}$ at atmospheric pressure, which are in the range of a few mm. The inclination affects the lower limit of the spatial resolution that can be obtained with the wires (sub-mm) but it does not affect the resolution given by the strips. This will be discussed in more details in Section~\ref{spatres}.
\\ The inclination improves the counting rate capability as well because a neutron flux is spread over a $11.5$ times wider surface ($1/\sin{\left(5^{\circ}\right)}=11.5$).
\\ A full characterization of the counting rate capability of MWPCs can be found in~\cite{RC_mathieson1,RC_mathieson2}. The counting rate capability of a MWPC is proportional to $1/\left(h^3\cdot s\right)$ with h the wire to cathode distance and s the wire pitch (both are about 4mm in the Multi-Blade). By changing h, the rate capability is strongly influenced~\cite{RC_andronic}. 
\\ Experimental studies on the response of MWPCs to the rate can be found in~\cite{RC_andronic,RC_Ivaniouchenkov,RC_koperny,RC_fonte}, all of them have been performed with X-rays and not neutrons. The high count rate does not trigger any breakdown in MWPC but only reduces the output amplitudes due to the accumulation of space charge~\cite{RC_Ivaniouchenkov}. In some high-rate measurements, the standard or thin-gap MWPC can be also be used if its spatial resolution satisfies the requirements~\cite{RC_fonte}.
\\ Almost all primary ion pairs are formed outside the multiplication region. After drifting towards the anode each electron produces, on average, the same number of secondary electrons by a charge multiplication process close to the anode wire. All secondary ion pairs generating the signal are thus approximately produced at the anode. Although the primary charge is not the same for X-rays and neutrons we can compare the two by normalizing to the same amount of secondary charge. When a neutron is converted in the $\mathrm{^{10}B_4C}$-layer, the neutron capture fragments that reach the gas create a primary charge of at most $9\,fC$ (alpha particle of $1470\,KeV$ in $\mathrm{Ar/CO_2}$ (80/20), $w=27eV$~\cite{DET_sauli1977}) which corresponds to about 55000 pairs. For a gas gain of $15$, we can expect at most $800000$ pairs ($0.1\,pC$). A $6.6\,KeV$ X-ray in the same gas mixture produces about $250$ primary pairs which corresponds to a gas gain of $3200$ to get the same secondary charge. Referring to~\cite{RC_andronic,RC_fonte} these MWPCs can reach counting rate capabilities beyond $10^4\,Hz/mm^2$ for the same geometrical parameters as the Multi-Blade. Hence due to the inclined geometry this number must be multiplied by a factor $11.5$. We expect then the Multi-Blade to show a counting rate capability beyond $10^5\,Hz/mm^2$. However, the feature of the primary charge has been omitted, thus experimental evidence is needed. 
\\ Note that the ion mobility differes for various gas types and it is generally constant with respect to the reduced electric field (E/P)~\cite{DET_ravazzani}. A larger pressure can be compensated by a larger electric field for a given geometry~\cite{RC_hendricks} . The large ions collection time is the main reason for the space charge effect and it can be of the order of ms. Note that a thicker wire can be used to speed up the ion collection time at the cathode by increasing the drift field while keeping the gas gain fixed, with the extra advantage of a more robust mechanical structure. One way to lower the space charge effect is to decrease the gas gain as much as possible (keeping a reasonable signal-to-noise ratio for the electronics) in order to diminish the total charge to evacuate per neutron converted. 
\subsubsection{Amplification alternatives}
In some cases the detector is operated in a vacuum tank, a requirement set by the instrument to avoid the scattering of the incoming neutrons. Since the Multi-Blade detector is operated at atmospheric pressure with a continuous gas flow, cost-effective materials can be used in the detector and the differential pressure on the detector entrance window is much smaller than that of $\mathrm{^3He}$-based detectors, as $\mathrm{^3He}$ detectors are usually filled with a few bars of gas. This implies that the thickness of the window of the Multi-Blade can be significantly reduced. In the event of atmospheric pressure in the neutron flight path, this means that a thin Al-foil can be used as the window.
\\ Replacing the wires with other readout systems is worth considering. It has been shown that Micro Pattern Gaseous Detectors (MPGD) can reach higher counting rate capability and good spatial resolution~\cite{MPGD_titov,MPGD_prospect,MPGD_shoji,MPGD_CrociRate}. It has been shown that Gas Electron Multipliers-based (GEM) neutron detectors can reach a few tens of $MHz/cm^2$~\cite{MPGD_GEMcroci,MPGD_CrociRate}. However, the inclined geometry of the Multi-Blade detector makes the choice of wires very natural. If for example we consider Gas Electron Multipliers (GEMs)~\cite{DET_GEM1}, those are made of kapton and even if they can be manufactured in very thin foils the inclination increases the effective thickness and results in excessively high scattering. 
In other applications~\cite{DET_doro1} where these layers are crossed perpendicularly by neutrons they represent a valid technology in neutron detection. 
\\Other possible MPGD choices are the Micro Strip Gas Chambers (MSGC)~\cite{DET_MSGC1,DET_MSGC2} ot the Micro-Mesh Gaseous Structure (MicroMegas)~\cite{DET_MicroMeg1,DET_MicroMeg2}, both of which present challenges when operating in this specific geometry of closely stacked cassettes. Moreover, MicroMegas employ PCBs which have scattering challenges for operation at an angle. 
\\ Moreover, the ionization chamber mode of operation (a gas gain of 1) of the wire chamber of the Multi-Blade is a possibility to reduce the space charge effect, but it is challenging for noise suppression. 
\subsubsection{Signal formation}
The Multi-Blade MWPC aims to work at a gas gain of $15$ and this is only possible if the anode wires and cathode strips are readout individually. This scheme reaches a larger signal-to-noise ratio with respect to the charge division readout~\cite{DET_chargediv}. In general, $\mathrm{^3He}$-based detectors use this readout and the secondary charge to evacuate is generally much larger: about $3\,fC$ (20000 pairs, $E=770KeV$, $w=37eV$~\cite{DET_sauli1977}) are created per neutron conversion and a gas gain of a few hundreds is needed to reach $1-2\,pC$ charge to get the required position resolution. Often they are operated in the region of limited proportionality in which the response is greatly affected by the space charge of a large amount of positive ions created in the avalanche; the effect of self-induced space charge~\cite{RC_koperny}. Note that the improvement in counting rate capability it is mainly connected the detector parameters chosen, including the readout scheme and only marginally to the isotope used to convert neutrons. 
\\At high rate operation, the individual readout (as opposed to charge division) is mandatory to disentangle hits occurring nearly at the same time (that is, unresolved due to the finite time resolution of the detector). The measured amplitudes on the wires and on the strips are strongly correlated (since they couple to the same avalanche), therefore with sufficient dynamic range, the ambiguity can be resolved most of the time by requiring matching amplitudes.
\section{Simulations}\label{simu}
\subsection{Cassette arrangement simulation}
\begin{figure}[htbp]
\centering
\includegraphics[width=.49\textwidth,keepaspectratio]{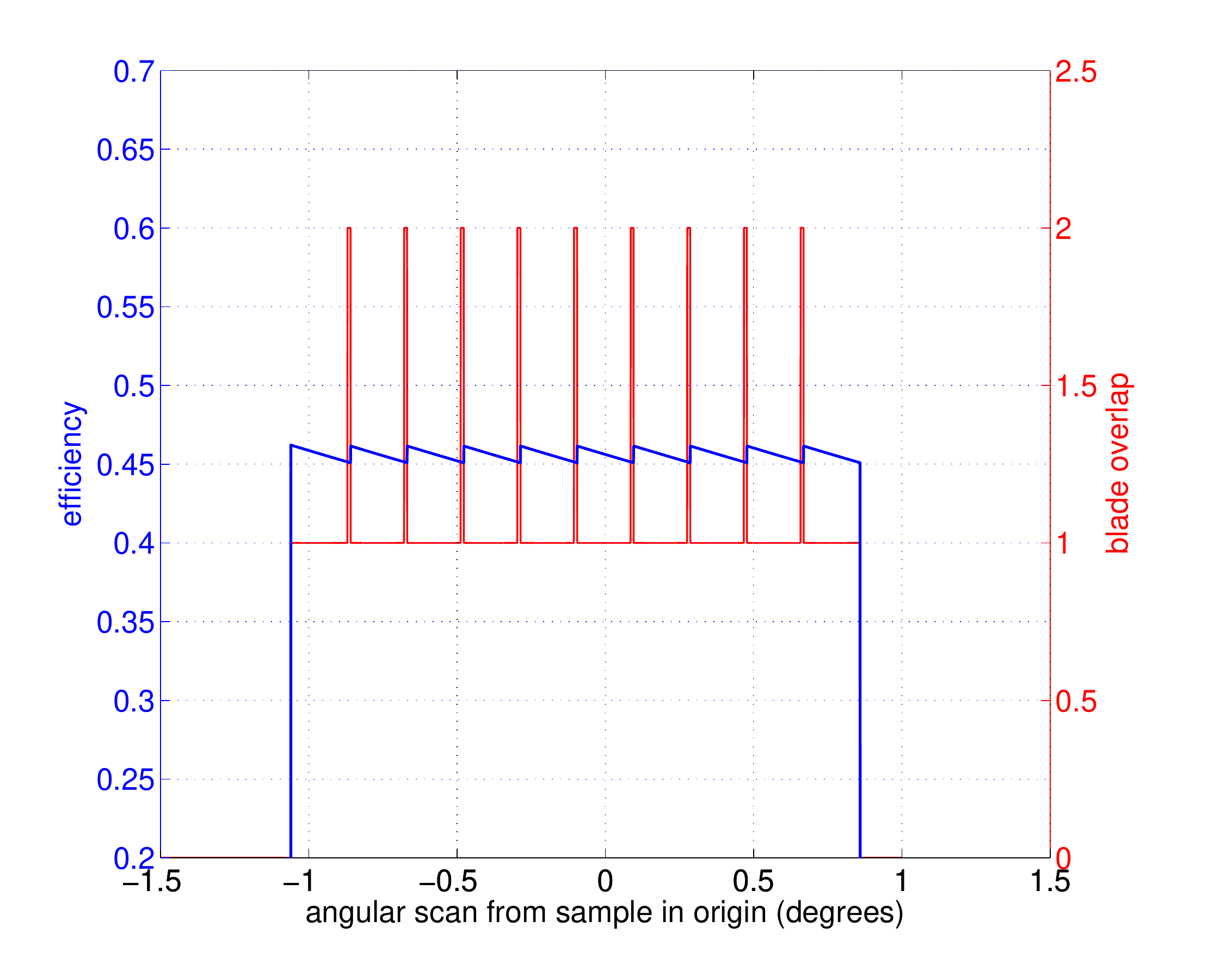}
\includegraphics[width=.49\textwidth,keepaspectratio]{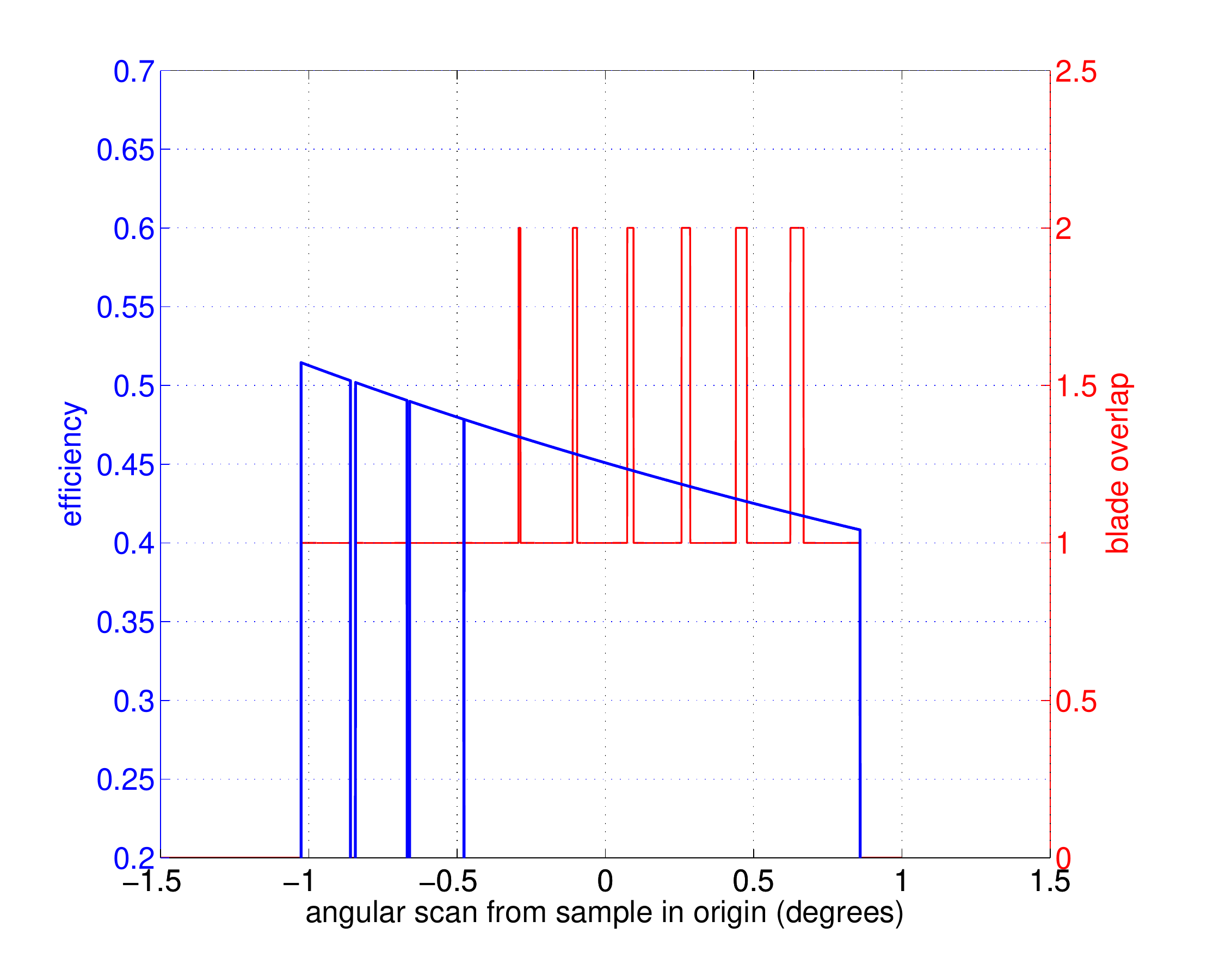}
\caption{\label{fig6}\footnotesize Calculated efficiency (blue) at $2.5$~\AA\,for $10$ cassettes arranged over a circle of radius $3\,m$ with sample in the center (left) and for 10 cassettes stacked parallel at $3\,m$ from the sample (right). In red is shown the overlap between the cassettes, this is only an integer number according upon how many cassettes are crossed by neutrons.}
\end{figure}

The Multi-Blade is a modular detector, a significant concern in a modular design is the uniformity of the response. Several effects might contribute to degrade the uniformity and they have to be taken into account in the detector concept. In particular the substrate flatness is an important aspect to keep the electric field uniform in the MWPC, however from the point of view of efficiency the overlap between different cassettes requires a carefully study. The cassettes must be arranged over a circle around the sample to provide a uniform response; the precision of the arrangement and the mechanical tolerances are the crucial aspects to consider from the mechanical point of view. 
\\ In the current detector the coated area of a substrate has been fixed to $X \times Y=130mm\cdot140mm$, thus at 5 degrees each blade subtends an area of about $11.3mm \,(=130mm \cdot \sin{\left(5^{\circ}\right)}) \times 140mm$. It is possible to increase the vertical dimension of the blade to match the area coverage required by the instruments, subject to mechanical considerations. For FREIA the Multi-Blade will have $300\,mm$-long blades placed horizontally and for ESTIA the cassettes will be placed vertically and they will be $250\,mm$ high. The number of blades needed to cover the other dimension is about one per cm.
\\The sample-to-detector distance is fixed for both reflectometers to $3\,m$ and $4\,m$ (Table~\ref{tab1}); and the neutron detection efficiency for a single $\mathrm{^{10}B_4C}$-layer ($>3\,\mu m$) inclined at 5 degrees is about $44\%$ at $2.5$~\AA~\cite{MIO_analyt}. We consider the sample to be a point-like source of neutrons. The specular incidence (and reflected) angle never exceeds a few degrees (i.e. $3$ degrees) and the sample surface has a physical extension of $80\,mm$ at most, then the projected sample dimension toward the detector is only about $4\,mm$ for a $3\,m$ distance. The difference between the angle of incidence of a neutron impinging the beginning and at the end of a $130\,mm$-blade at a distance of $3\,m$ is comprised within $0.2$ degrees. The efficiency then only varies within $1\%$ due to the change in angle, while a wider substrate will cause a larger efficiency variation. Figure~\ref{fig6} (left) shows the calculated efficiency~\cite{MIO_analyt} for 10 cassettes arranged circularly around a center of scattering, i.e. the sample, at $3\,m$ distance. Each cassette shows the expected variation of efficiency within $1\%$. On the other hand, if the cassettes are stacked parallel to each other (right plot in Figure~\ref{fig6}), the variation of the efficiency is still about $1\%$ within each cassette but it varies dramatically from one to another. This variation is about $10\%$ from the first to the last unit. It is then crucial to arrange the cassettes over a circle around the sample in order to keep the variation of the efficiency within $1\%$. 
\\The cassettes are arranged to have some overlap; i.e. each blade makes a shadow over the adjacent in order to avoid dead areas. In this set-up approximately $15\,mm$ of the active region of each cassette is shadowed by the adjacent unit, which corresponds to $1.3\,mm$ at 5 degrees. Therefore, if each cassette subtends an area of $11.3\,mm \times 140\,mm$, this area is reduced to $10\, mm\times 140\,mm$ due to the shadowing effect. Referring to Figure~\ref{fig6}, the overlap between units is shown in red. This function assumes only integer values according to the number of blades crossed by a neutron path from the sample to the detector. The efficiency is only plotted for one or no converters crossed and any further layer above one does not contribute to the efficiency because the layers is supposed to absorb all the neutrons. 
\\A more extended substrate may help to relax the requirement on the overall thickness of a cassette, presently $10\,mm$. A blade of $X=130\,mm$ represents already a challenge in terms of mechanical available space. Increasing $X$ results in a reduced number of cassettes to cover a given area, but a too large extension of the blade will degrade the uniformity of the detector because of the variation of the incidence angle. 
\\ A misalignment of the detector angle affects overall the response more than a misalignment in the distance to the sample. E.g. a variation of the arrangement angle of $\pm0.5$ degrees causes a variation of the efficiency of about $4\%$ whereas a mispositioned detector of $\pm 0.2m$ only causes a variation in efficiency of approximately $2\%$.
\subsection{Electric field simulation}\label{esim}
In the current detector the units are arranged over a circle of $3\,m$ radius and the relative angle between two adjacent cassettes is $0.19$ degrees. The wire plane and the converter ($\mathrm{^{10}B_4C}$-layer) are parallel and at $4.6\,mm$ distance. They physically belong to the same cassette. The strip plane of each MWPC belongs instead to the adjacent unit which is positioned with $+0.19$ degrees angle, i.e. the strip plane is not parallel to the wire and converter planes. This geometry affects the electric field and it is the price one has to pay to remove any material from the path of the incoming neutrons that can cause scattering. The wire pitch is $4\,mm$ and they have a diameter of $15\,\mu m$. The gas mixture is $\mathrm{Ar/CO_2}$ (80/20) continuously flushed at atmospheric pressure. The wires are labelled from $1$ to $32$ as shown in Figure~\ref{fig1}. 
\begin{figure}[htbp]
\centering
\includegraphics[width=.49\textwidth,keepaspectratio]{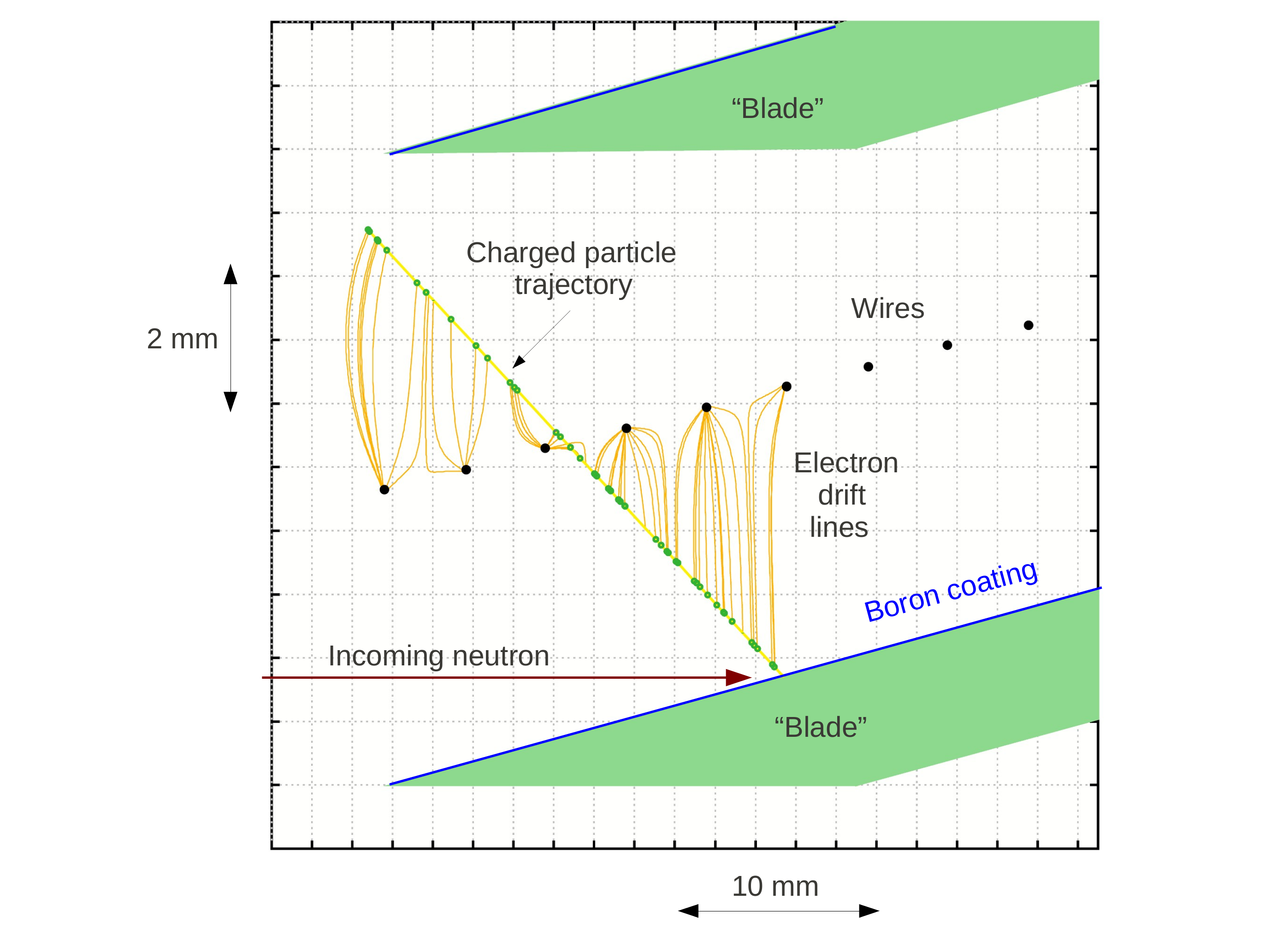}
\includegraphics[width=.49\textwidth,keepaspectratio]{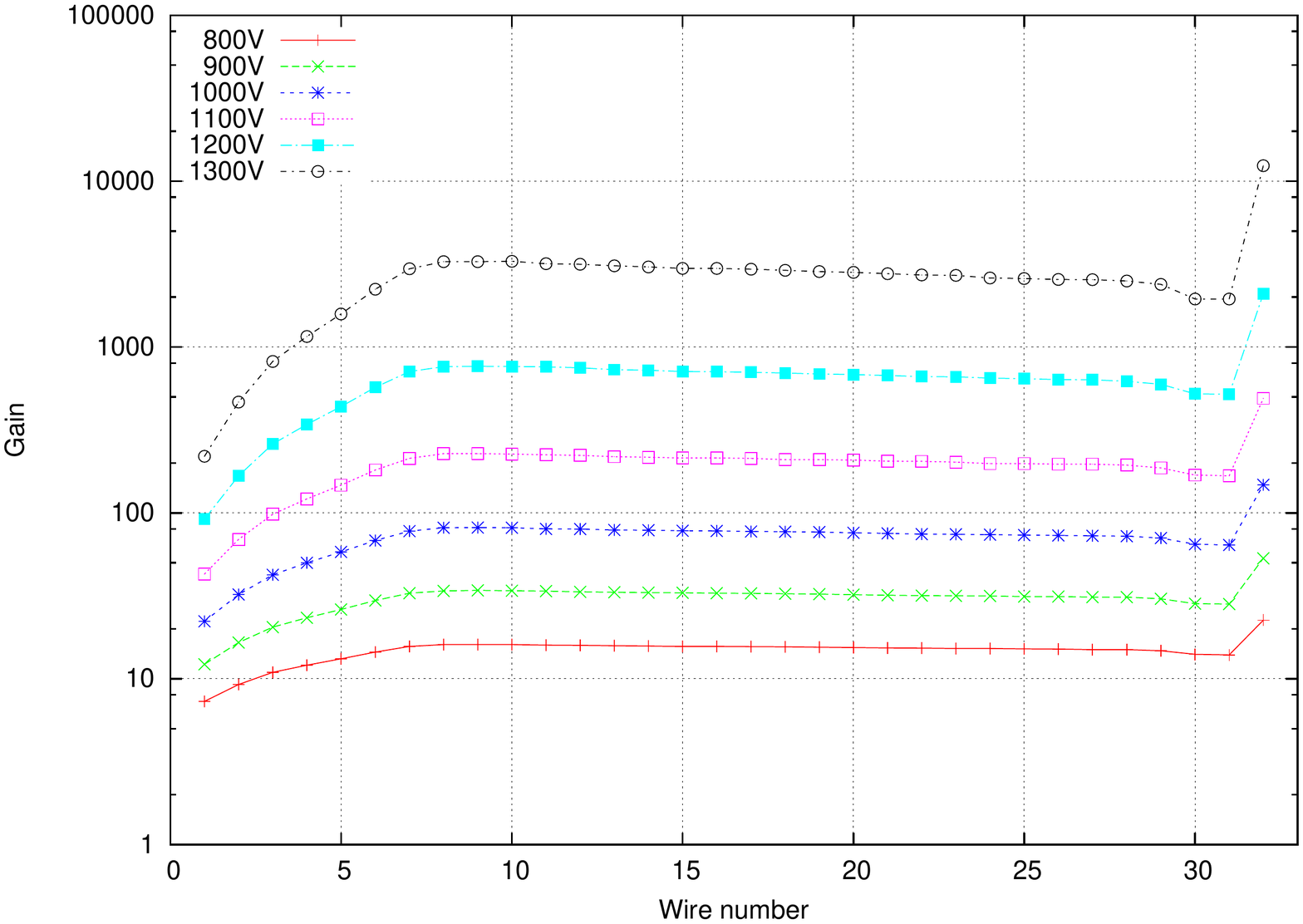}
\caption{\label{fig7} \footnotesize Sketch of the simulated process of a neutron converted in the boron-layer and a charge particle escaping from the layer generating electrons along the trajectory and drifting according to the field lines (left). Gas gain per wire for several applied potentials (right).}
\end{figure} 
\begin{figure}[htbp]
\centering
\includegraphics[width=.49\textwidth,keepaspectratio]{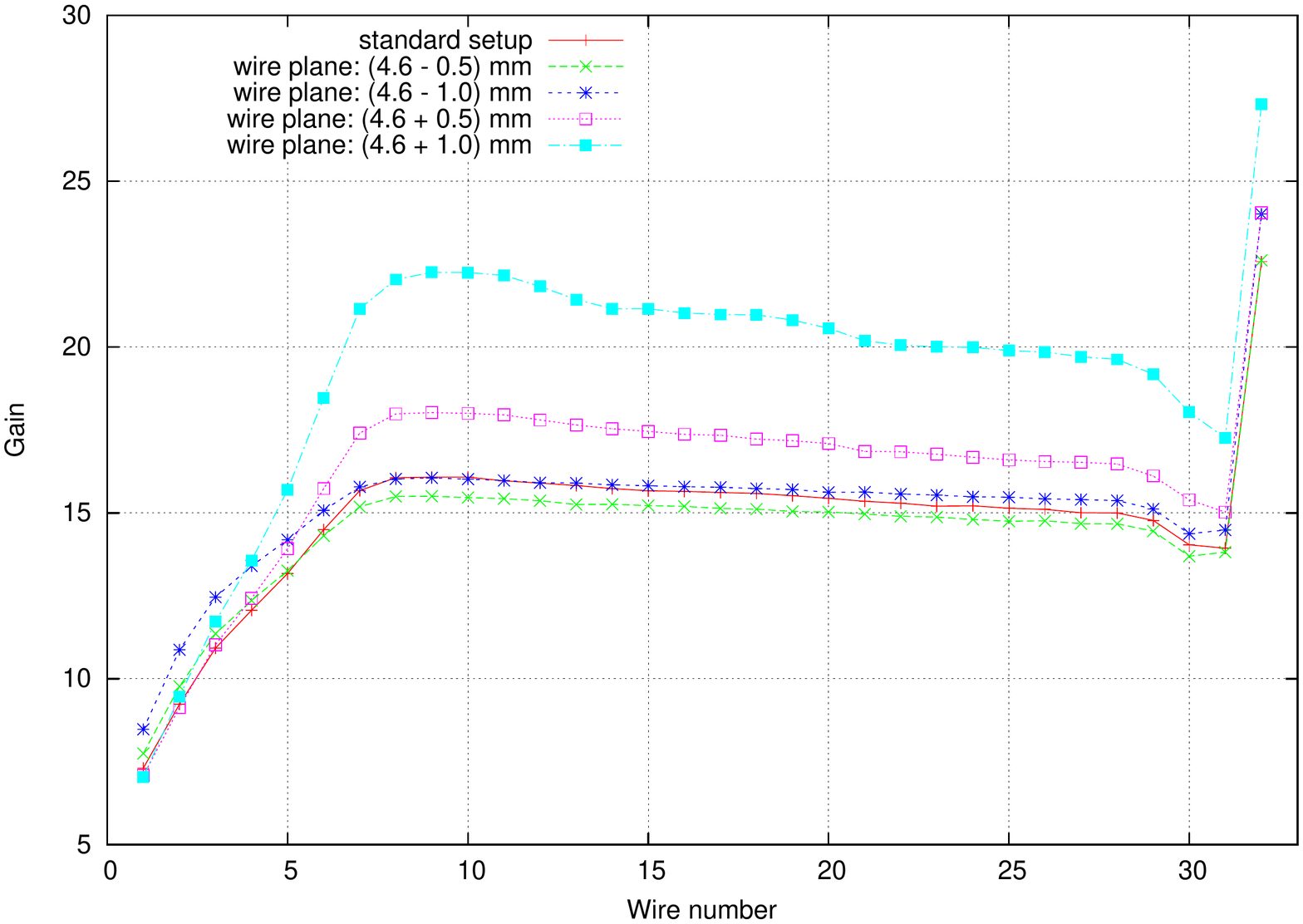}
\includegraphics[width=.49\textwidth,keepaspectratio]{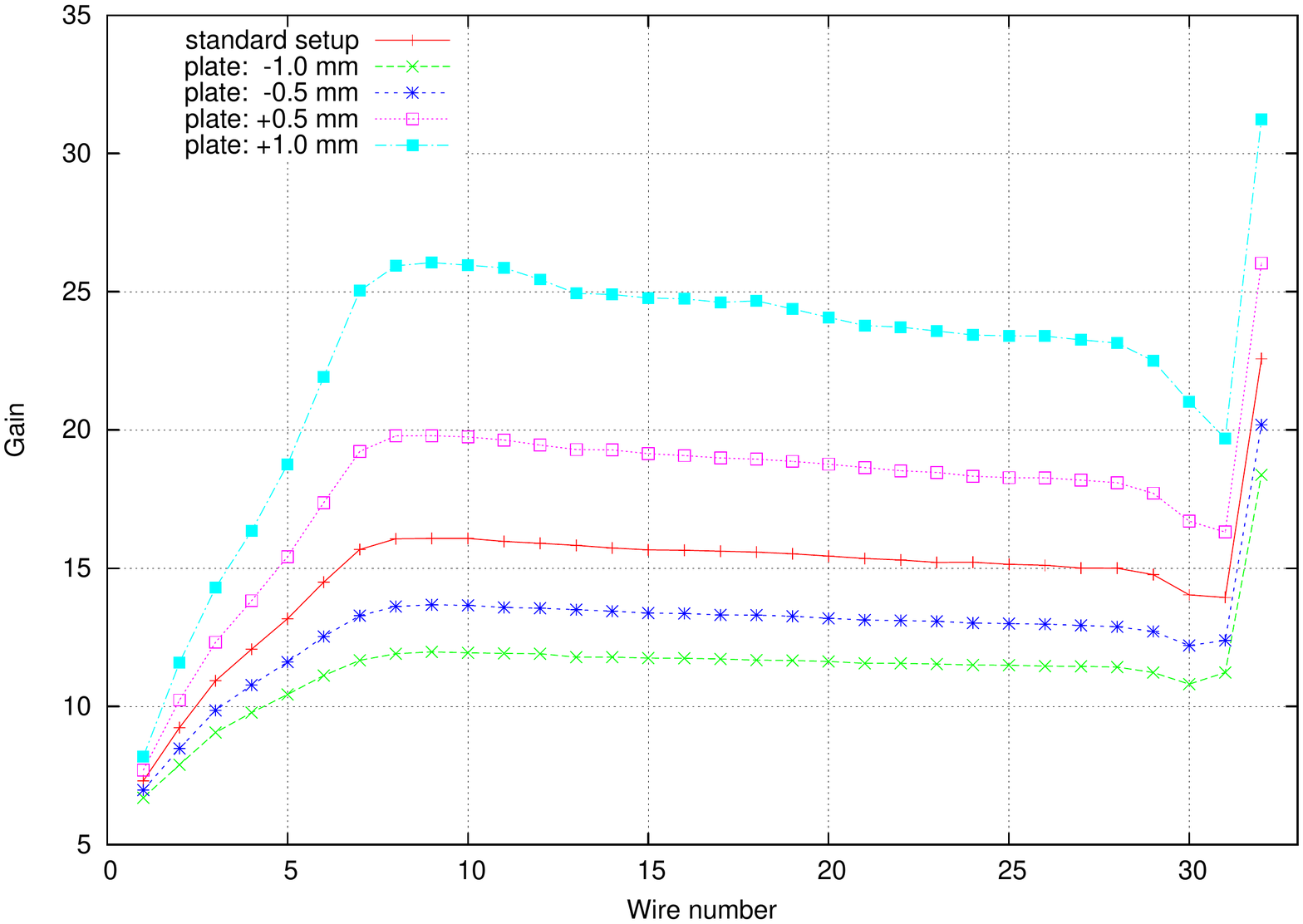}
\caption{\label{fig8} \footnotesize Gas gain per wire for a given voltage of 800V for several position of the wire plane while keeping the strip plane fixed (left) and for several position of the strip plane while keeping the wire plane fixed (right). }
\end{figure} 
Figure~\ref{fig7} (left) shows a sketch of the simulated process of a neutron converted by the boron-layer and a charged particle escaping from the layer generating electrons along the trajectory and drifting according to the electric field lines. For each electron we calculate the integral of the Townsend coefficient over the drift path. This represents the mean number of secondary electrons produced along the trajectory of the electron that is drifting, assuming exponential avalanche fluctuations and neglecting attachment. The gas gain given per wire is shown on the right in Figure~\ref{fig7} for several voltages applied to the wires. At a voltage of $800\,V$ we expect a gas gain of $15$. The gas gain varies a lot along the first 7 wires due to the increase in the cathode gap while moving toward the edge of the cassette. There is a slight reduction of the gain from wire 8 until the end of the units and it is due to the relative angle between the cassettes (0.19 degrees). The variation in gain is smaller for lower voltages. 
\\ Figure~\ref{fig8} (left) shows the gas gain given per wire for a voltage of $800\,V$; the converter and the opposite strip plane are kept fixed while the wire plane is shifted orthogonally to the converter plane. A positive shift means that the wire plane and the strips are closer, while a negative shift indicates that the wire plane is moving closer to the converter. On the right, in Figure~\ref{fig8}, the wire plane and converter are kept at a fixed position and the strip plane is shifting. The first set-up corresponds to a misalignment of the wire plane within a single cassette, whereas the second set-up corresponds to a misalignment in the arrangement of the cassettes. We can conclude that a misalignment in the wire plane positioning is less crucial than a misalignment in the cassette arrangement. A $0.5\,mm$ deviation for the wire positioning can be tolerated, and corrected by threshold adjustment; the constraint is much more strict for the arrangement of the cassettes. 
\\As a possible solution, in order to compensate the reduction in gain and to improve the dynamic range, the wire thicknesses or pitch for the first 7 wires can be adjusted accordingly to the gain variation. Otherwise the gain drop at the first 7 wires can be compensated with a separate high voltage supply or by adjusting individual threshold, in hardware or software, on each channel. 
\section{Experimental setup}
\begin{figure}[htbp]
\centering
\includegraphics[width=.7\textwidth,keepaspectratio]{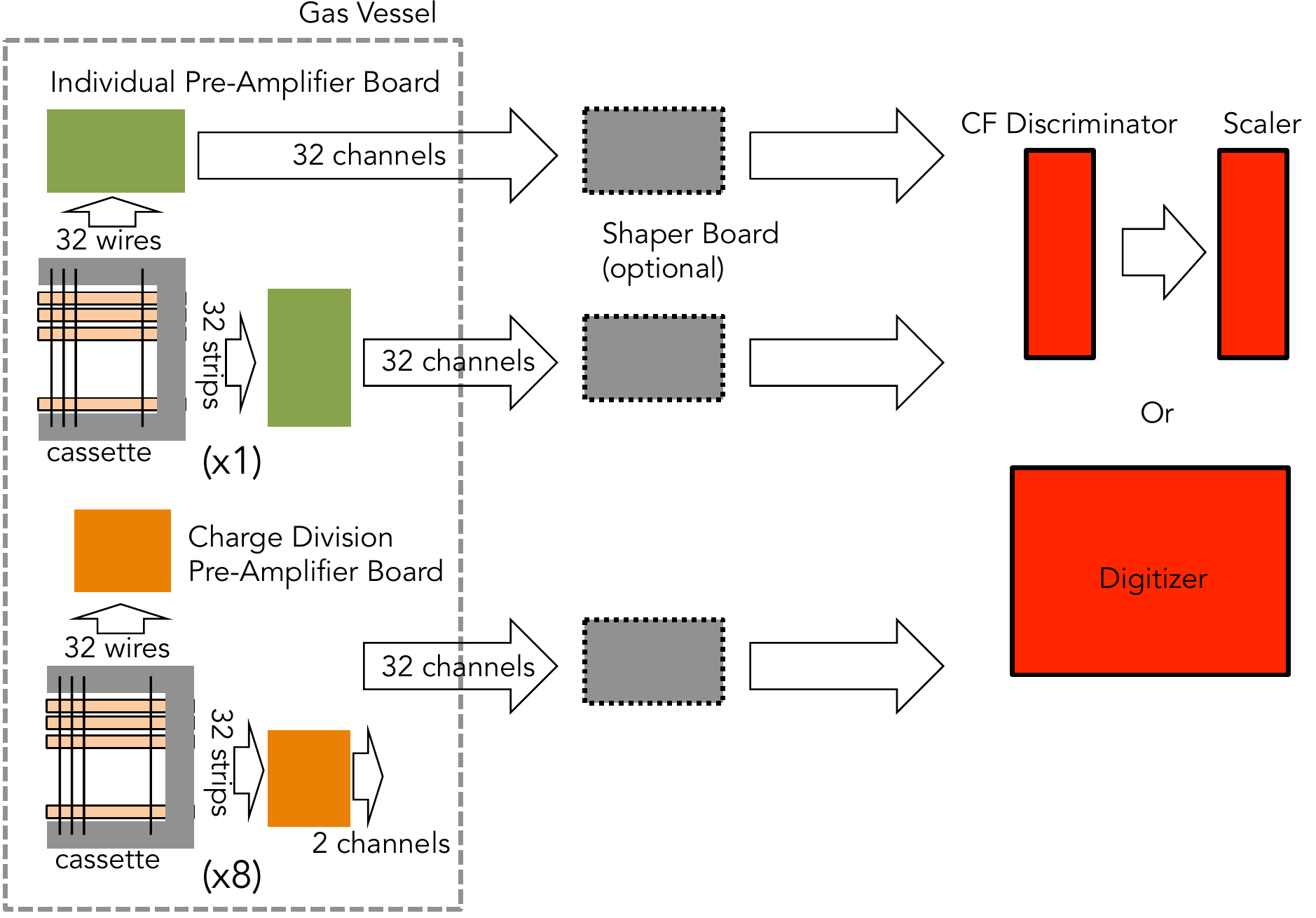}
\caption{\label{figex1} \footnotesize A sketch of the electronics readout chain.}
\end{figure}

The tests were performed at the triple axis spectrometer ATHOS at BNC~\cite{FAC_BNC} (Budapest - Hungary). The preliminary test of the Multi-Blade and the $\gamma$-sensitivity measurements were performed at the Source Testing Facility (STF)~\cite{SF1} at the Lund University (Sweden). 
\\The demonstrator is made up of 9 cassettes, each one equipped with 32 wires and 32 strips. The readout of those channels is performed depending on the type of pre-amplifier board is plugged in. Two possible readouts are possible: a charge division readout (as it was for the previous demonstrator~\cite{MIO_MB2014}, Multi-Blade 2013) that allows to reduce the number of channels to 4 per cassette and an individual readout option with 64 channels per cassette (one amplifier for each wire or strip). In the tests one cassette was equipped with an individual readout (64 channels) and the remaining 8 cassettes with the charge division readout, and thus 32 channels in total. The presented configuration allowed to keep the number of readout channels reasonably low by giving the possibility to test a complete detector. The detector has been tested at $800\,V$ for the individual readout (gas gain $\approx 15$) whereas a voltage of $1300\,V$ (gas gain $\approx 2000$) has been used for the charge division readout in order to get a suitable signal-to-noise ratio and the corresponding spatial resolution. In the final configuration for a high rate detector, the gas gain must be kept as low as possible in order to reduce the space charge effects and to exploit the maximum counting rate capability, thus the individual readout is the sole option. Figure~\ref{figex1} shows a sketch of the readout electronics chain. 
\\ The pre-amplifiers are CREMAT~\cite{EL_cremat} $CR-110$ charge sensitive pre-amplifiers and are placed inside the gas vessel; any signal can be optionally shaped with a CREMAT~\cite{EL_cremat} $CR-200-500ns$ Gaussian shaping amplifier or directly sent to the acquisition system. This means that we may choose to record a pre-amplifier signal or a shaped signal. The pre-amplifiers have a gain of $1.4\,V/pC$ and and the shaping amplifiers a gain of $10$ with $500\,ns$ shaping time. A CAEN~\cite{EL_CAEN} Digitizer is used to record the traces (Mod. DT5740, $12\,bit$, $62.5\,MS/s$).
\\In order to quantify the actual counting rate capability of the detector itself, for the specific counting rate measurement we use a discriminator and a scaler bypassing the digitizer. The scaler was a CAEN~\cite{EL_CAEN} V830 with maximum counting rate of $250\,MHz$ and the constant fraction discriminator was a CAEN~\cite{EL_CAEN} V812.
\section{Results}
\subsection{Signals, PHS and gas gain}\label{phchapt}
\begin{figure}[htbp]
\centering
\includegraphics[width=.49\textwidth,keepaspectratio]{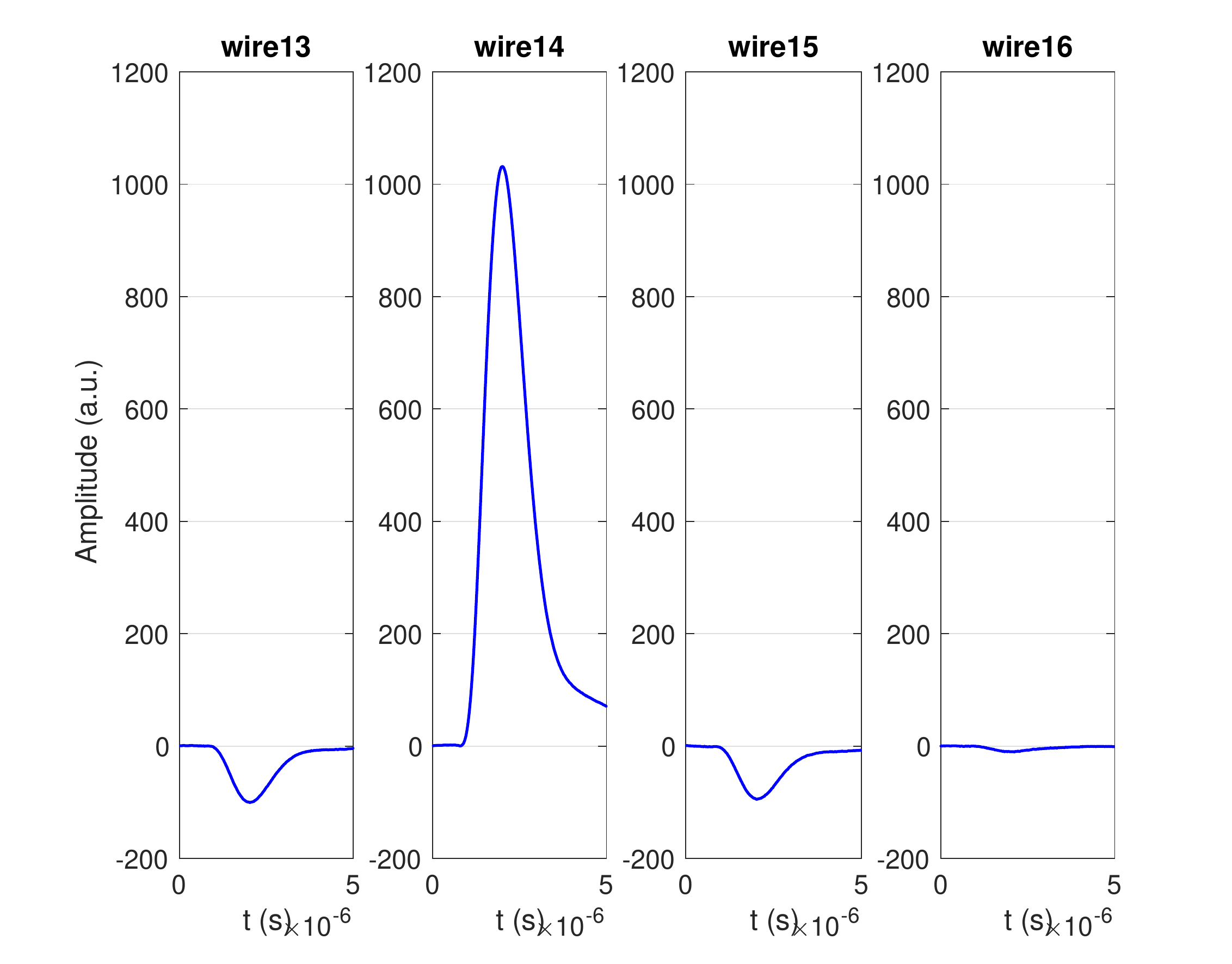}
\includegraphics[width=.49\textwidth,keepaspectratio]{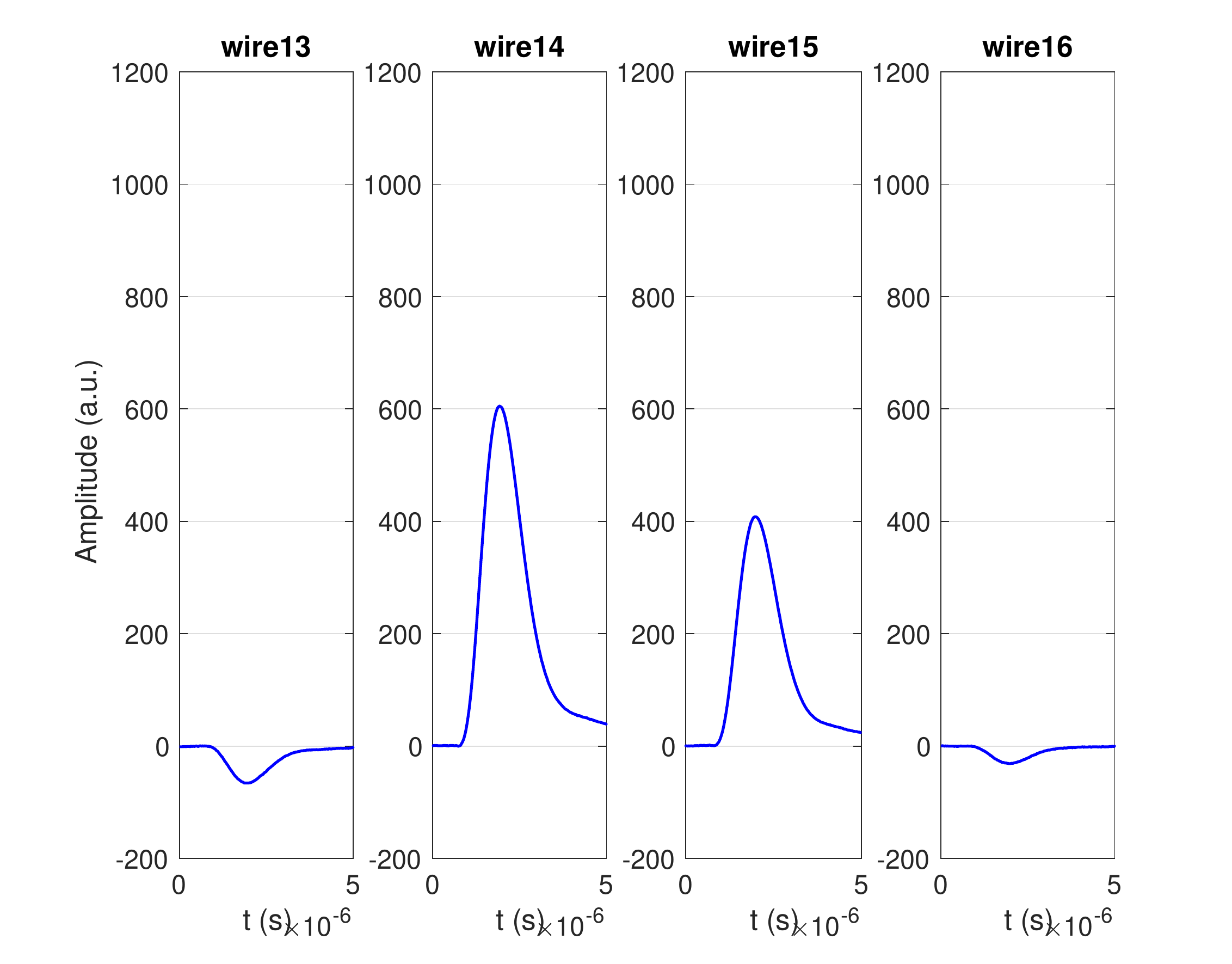}
\caption{\label{fig70} \footnotesize The signals (after shaper) of four adjacent wires of the cassette read out with individual amplifiers. A neutron event generates a charge corresponding to $\approx 1000$ in ADC levels in both plots. Depending on the particle (from the neutron capture reaction) escape orientation from the layer, one wire (left) or two wires (right) can be involved in the detection process.}
\end{figure}

\begin{figure}[htbp]
\centering
\includegraphics[width=.7\textwidth,keepaspectratio]{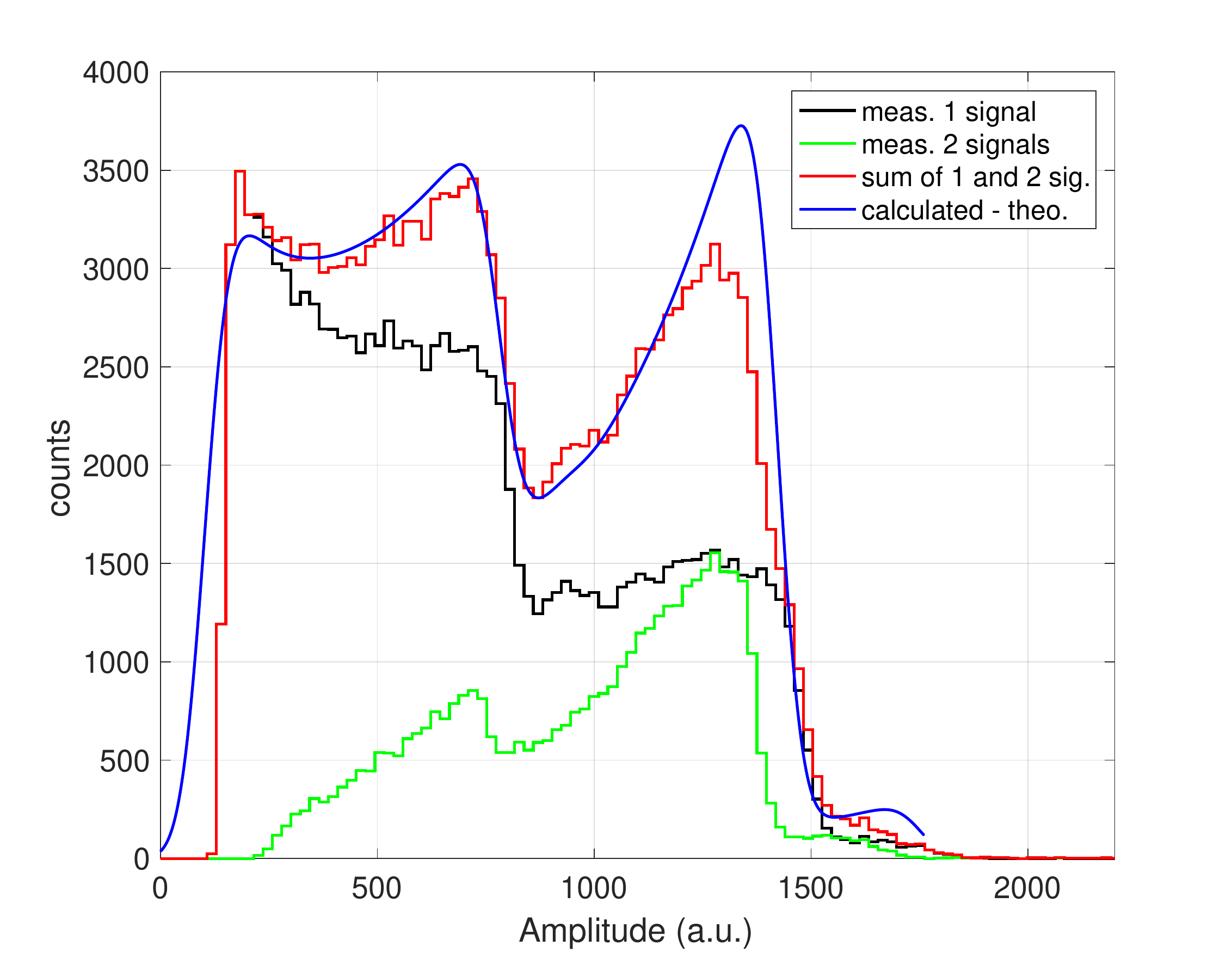}
\caption{\label{fig71} \footnotesize Calculated (blue) and measured PHS at $800\,V$. The measured PHS when only one wire (black) or two wires (green) are involved in the detection process and the total (red).}
\end{figure}

When a neutron is converted, the capture reaction fragments can escape the layer with any orientation and due to the random energy loss, the deposited charge in the gas volume ranges from zero to the full energy carried by the particle. The alpha particle ($1470\,KeV$) maximum range in $\mathrm{Ar/CO_2}$ at atmospheric pressure is about $8\,mm$, and this corresponds to twice the wire (strip) pitch. Figure~\ref{fig70} shows the signals (after the shaper) collected at four adjacent wires. These wires are in the center of the cassette (wires from 13 to 16 out of 32) read out with individual amplifiers. In both plots the full energy carried by the particle is the same and corresponds to $\approx 1000$ in ADC levels. In the plot on the left the total charge is collected by one wire whereas in the plot on the right it is collected by two wires demonstrating the orientation effect. Moreover, when the ion charge created in the avalanche close to the wires travels toward the cathodes it induces a signal on the involved wire and a negative pulse is also induced on the neighbouring wires not directly involved in the process. 
Due to the geometry of the MWPC of the cassette about $70\%$ of the times a single wire is involved in the detection process, about $30\%$ two wires are firing; the probability to get three or more wires involved in a detection process is below $1\%$. 
\\ The capacitive signal coupling towards the strips is different from that of wires and it implies that most of the time two strips are involved in a detection process. About $25\%$ of the times only one strip or three strips are involved and $50\%$ of the times two strips are firing, and the probability to get four or more strips firing is below $1\%$. 
\\ Figure~\ref{fig71} shows the measured PHS at $800\,V$ on an individual wire and its neighbours. The black curve is obtained by adding together all the events that are collected at that wire (when a single wire is fired for each neutron event); the green curve is obtained by summing the amplitudes of that wire and the right or left neighbour, i.e. events that involve two wires. The red curve is the sum of the two measured PHS and the blue curve is the calcualted PHS obtained as shown in~\cite{MIO_analyt} with an energy resolution of $50\,KeV$ and an energy threshold of $100\,KeV$. 
\\ Based on the feature of the PHS, knowing the amplifier gain ($14\,V/pC$) and that the alpha particle generates at most on average $9\,fC$ of primary charge, we extrapolate a gas gain of $(20\pm3)$. 
\\ From the simulations in Section~\ref{esim} we expect the gas gain to be approximately $15$ and that a deviation of $0.5\,mm$ in the positioning of the cathode plate produces an error of $\pm 3$ on the gas gain. This is compatible with the experimental evidences discussed in this section~\footnote{As already mentioned in Section~\ref{sect2} in the present demonstrator the converter substrates are made of aluminium which suffers from deformation in the coating process.}.
\subsection{Efficiency}\label{effsusect}
\begin{figure}[htbp]
\centering
\includegraphics[width=.7\textwidth,keepaspectratio]{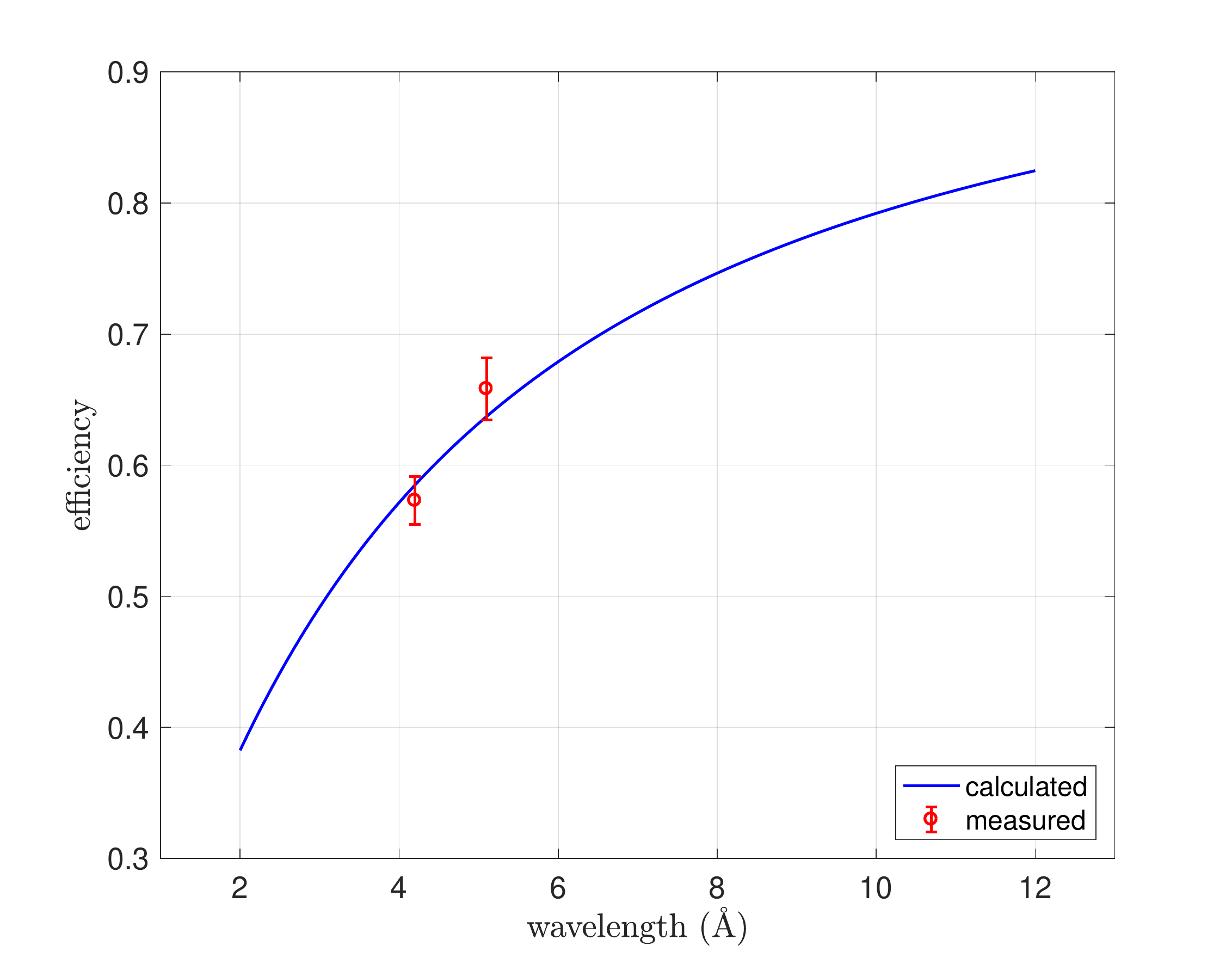}
\caption{\label{fig72} \footnotesize Measured and calculated efficiency with a $100\,KeV$ energy threshold.}
\end{figure}

It has been shown in~\cite{MIO_MB2014} that the theoretical efficiency can be reached at an inclination of $10$ degrees of the converter layer but it was not possible to prove this agreement at $5$ degrees. Moreover, it has been shown in~\cite{MIO_B10refl}  that the inclination of $5$ degrees does not affect the neutron detection efficiency due to neutron reflection at the layer surface. The neutron detection efficiency of the Multi-Blade detector at $5$ degrees has been measured.
\\The neutron beam was collimated with two collimation slits corresponding  to a $2\times10\,mm^2$ footprint on the detector and a PHS was recorded for $200\,s$. The integral of the PHS above the threshold and normalized to the incoming flux gives the efficiency. The measurement was repeated to calibrate the neutron flux with a $\mathrm{^3He}$-based detector filled with $8\,bar$ $\mathrm{^3He}$ and $1\,bar$ $\mathrm{CF_4}$ which is $3\,cm$ thick. We assume this detector to be fully efficient at the wavelength we measure the flux and that between the measurement with the Multi-Blade and the $\mathrm{^3He}$ detector the reactor does not fluctuate. This measurement was performed for two neutron wavelengths: 4.2 and 5.1\AA, resulting in $\left( 56 \pm 2\right)\%$ and $\left( 65 \pm 2\right)\%$ respectively.
\\ Figure~\ref{fig72} shows the measured and calculated efficiencies according to~\cite{MIO_analyt} with a $100\,KeV$ energy threshold. 
\subsection{Overlap and Uniformity}
\begin{figure}[htbp]
\centering
\includegraphics[width=.38\textwidth,keepaspectratio]{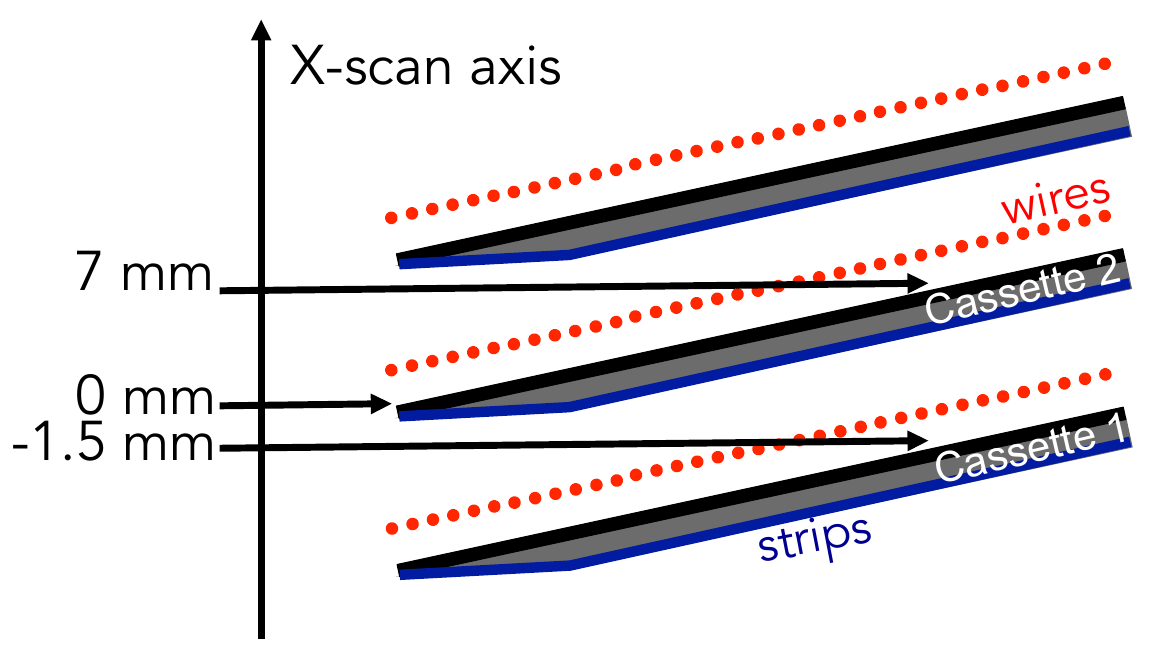}
\includegraphics[width=.60\textwidth,keepaspectratio]{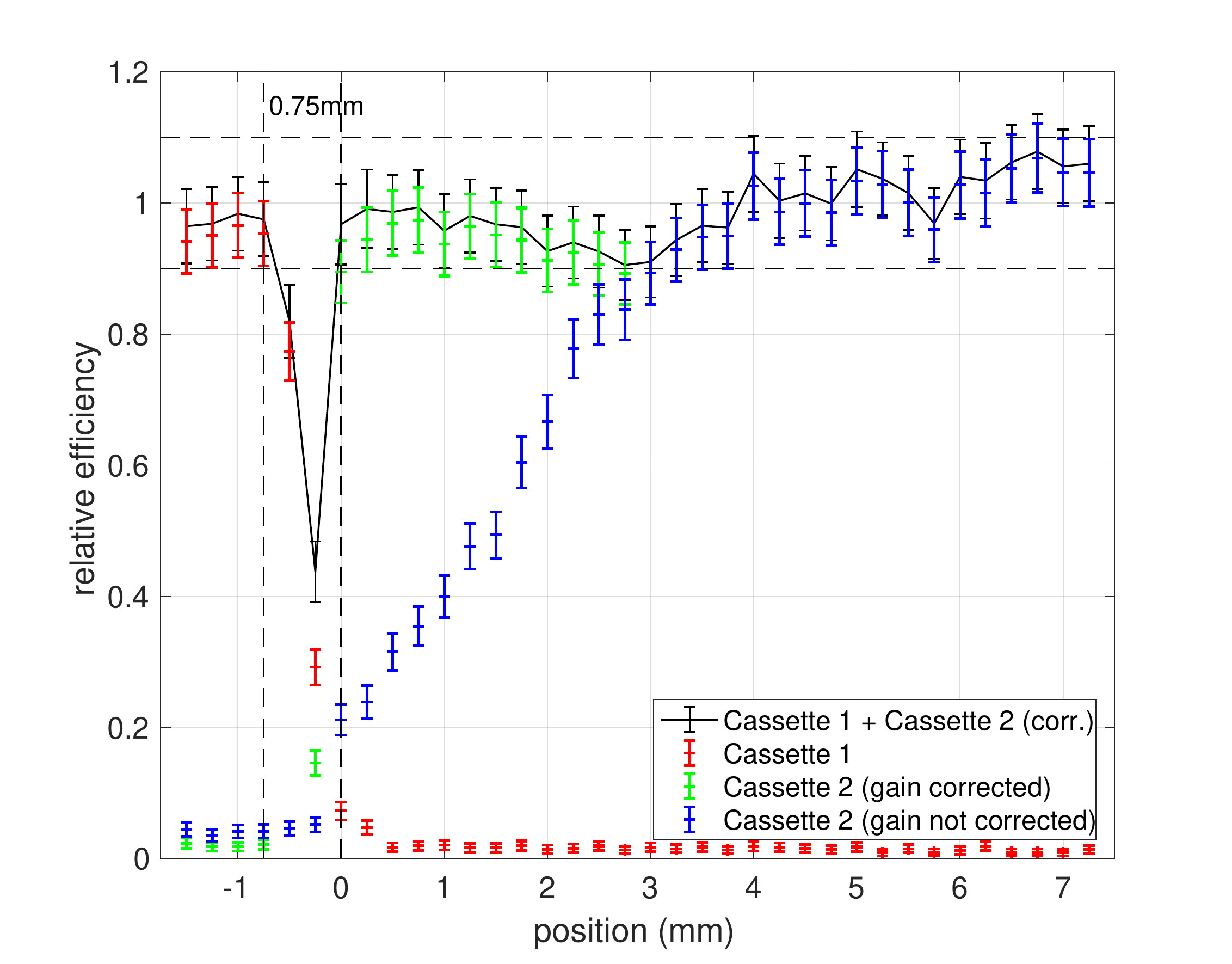}
\caption{\label{fig73} \footnotesize A drawing of the scan setup (left). The relative efficiency scan across the wires of 2 cassettes with a $0.25\,mm$ step (right). In blue the relative counts in the PHS as plotted as is for a fixed threshold whereas in green the threshold is modified according to the gain variation expected for the frontal part of the cassette. The two dashed horizontal lines represent a $20\%$ variation band.}
\end{figure}

The neutron beam was collimated by using two collimation slits to a about $0.2\times10\,mm^2$. The detector was scanned across the wires of two adjacent cassettes read out by charge division with a step of $0.25\,mm$. A PHS was recorded for each cassette and for each position of the scan. The number of counts in the PHS is plotted in Figure~\ref{fig73} as a function of the position and normalized to the average counts, i.e. the relative efficiency is plotted as a function of the position.
The zero position is aligned with the edge of the cassette 2. As shown in Section~\ref{esim} we expect the gas gain to vary in the first part of the cassette due to the geometry, this is shown by the blue experimental points that have been obtained by keeping the threshold fixed for any position of the scan. If the threshold is changed according to the change in gain we observe in the PHS, we obtain the green curve. At the cassette edge, in a range of $0.75\,mm$ the detector is fully efficient and two experimental points lay in this distance. The relative efficiency drops to $80\%$ and to slightly less than $50\%$ for these two points. The edges of the two binned points result in a total gap of $0.5\,mm$. In the prototype presented in~\cite{MIO_MB2014}, the Multi-Blade 2013, this gap was about $2\,mm$. 
\\ This measurement was performed at $1300\,V$, suitable for the charge division readout, and the overall variation of gain across a cassette is within $20\%$, the measurement was repeated on another cassette and the same variation was found. We also repeated this scan over the single cassette equipped with the individual readout and we found a variation within $10\%$. We expected the uniformity to improve for a lower gas gain (used in the individual readout) because the electric field is weaker and it tolerates larger mechanical imperfection. 
\subsection{Spatial Resolution}\label{spatres}
The spatial resolution of the detector was measured in both directions by scanning the surface with a collimated beam of about $0.2\times10\,mm^2$ size. The definition adopted for the resolution across the wires is based on the Shannon information theory and given in~\cite{DET_patrickinfo} was already used in~\cite{MIO_MB2014}. Note that the resolution of this detector is not primarly determined only by the wire or strip pitch since the track length of the neutron capture reaction fragments is comparable to the wire/strip pitch. In~\cite{MIO_MB2014} was shown that the resolution on wires is not a single wire pitch neither two, but rather something in between. 
\\ The measurement was performed by using both the charge division readout and the individual readout. The position of an event can be reconstructed by assigning triggered event to the wire or strip with with the largest amplitude (ADC level) regardless of how many wires or strip are involved in the detection process. Otherwise a Center of Gravity (CoG) algorithm can be used; it weights the different amplitudes of neighbour wires (or strips) to get a finer spatial resolution. With the charge division readout the CoG reconstruction is performed by default and the only way to compare the two algorithms is with the individual readout. 
\begin{figure}[htbp]
\centering
\includegraphics[width=.95\textwidth,keepaspectratio]{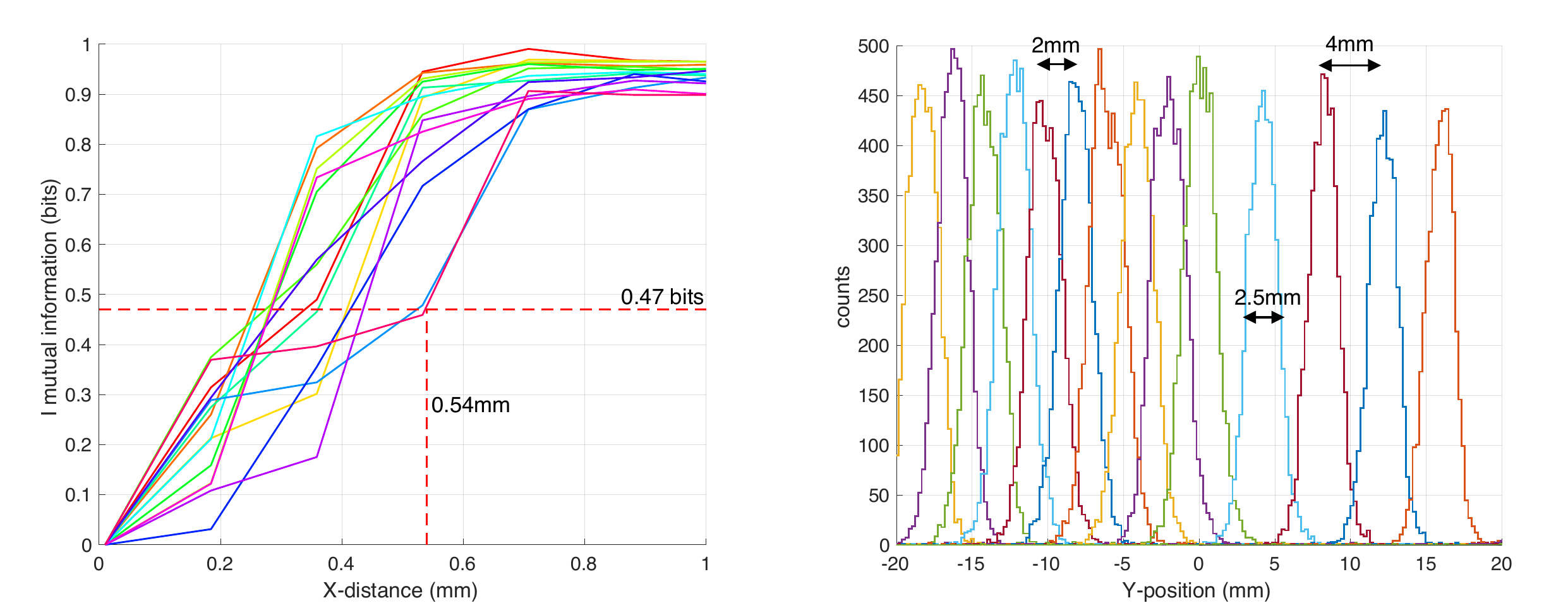}
\caption{\label{fig7499} \footnotesize The spatial resolution across the wires (left) according to the definition given in~\cite{DET_patrickinfo} and across the strips (right). Both are calculated by using the CoG algorithm.}
\end{figure} 
By using the largest amplitude reconstruction algorithm we get $0.59\,mm$ spatial resolution across the wires (X) and if we use the CoG algorithm (for both individual and charge division readouts) the resolution is modestly improved and it is $0.54\,mm$ according to the definition given in~\cite{DET_patrickinfo} (see Figure~\ref{fig7499}).
\\ From the scan on the strips the resolution is approximately the strip pitch if the largest amplitude algorithm is used (a similar result was already found in~\cite{MIO_MB2014}); but it can be improved to $2.5\,mm$ by using the CoG algorithm (see Figure~\ref{fig7499}). The fact that the resolution across the strips improves to a larger extent with respect to that of wires, is due to the fact that the probability for an event to involve more than one strip is larger for strip than that of wires (see Section~\ref{phchapt}).
\subsection{Stability}
The Multi-Blade was flushed with $\mathrm{Ar/CO_2}$ (80/20, fractions by volume) and a full volume ($\approx 30\,l$) of the detector was renewed every day (24h). A PHS was recorded every 10 minutes during a night with the neutron collimated beam impinging on the cassette read out with individual amplifiers. A variation of counts (for a fixed threshold) within $1\%$ was found over $12\,h$. The measurement was repeated and it gave similar results. 
\\ We expect the variation in counts to follow the atmospheric pressure variation if either the HV or threshold or the pressure in the vessel are not adjusted accordingly; however a stability within $1\%$ is sufficient to operate the detector for most of the neutron reflectometry measurements. 
\subsection{Counting rate capability and shaping time}
\begin{figure}[htbp]
\centering
\includegraphics[width=.49\textwidth,keepaspectratio]{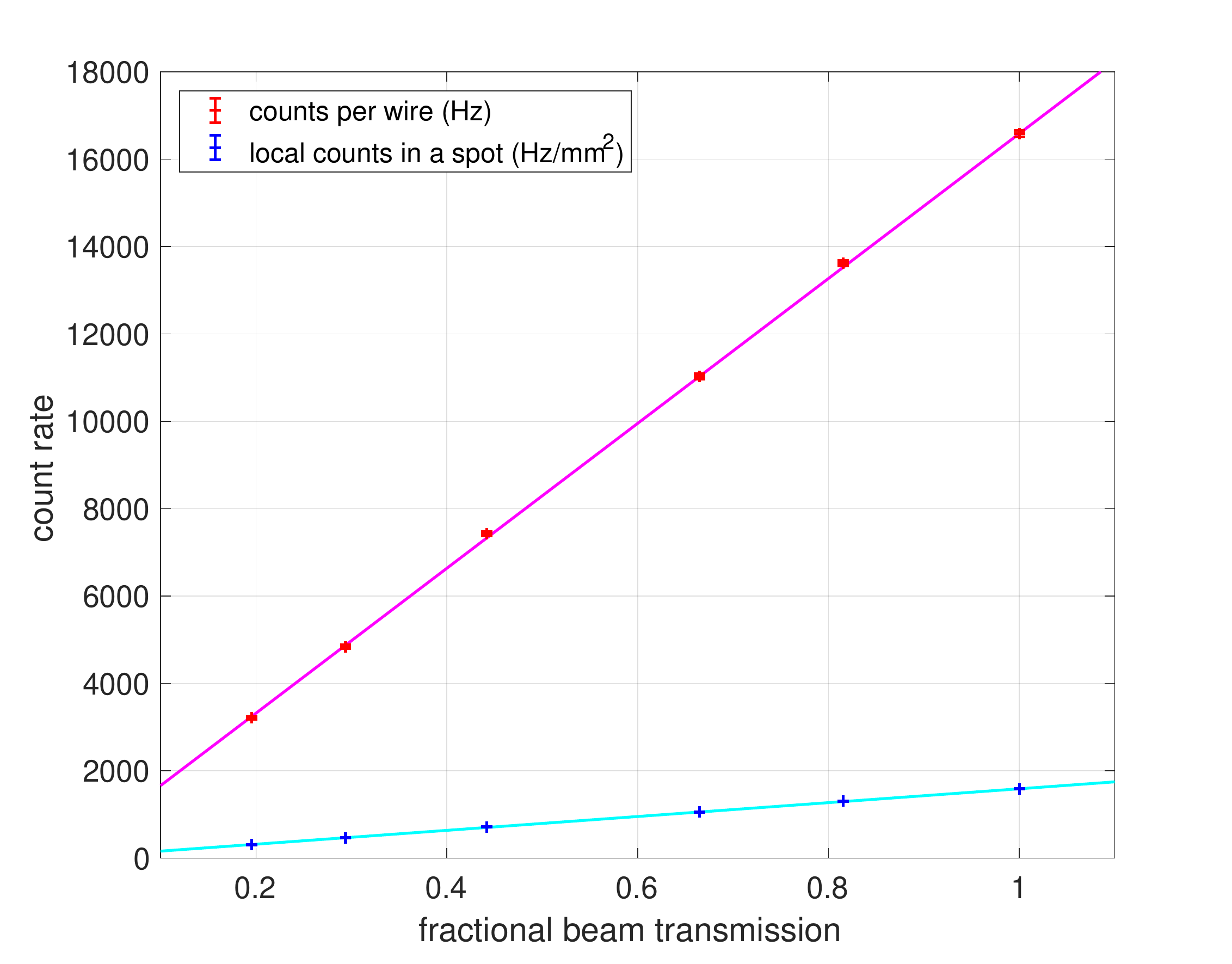}
\includegraphics[width=.49\textwidth,keepaspectratio]{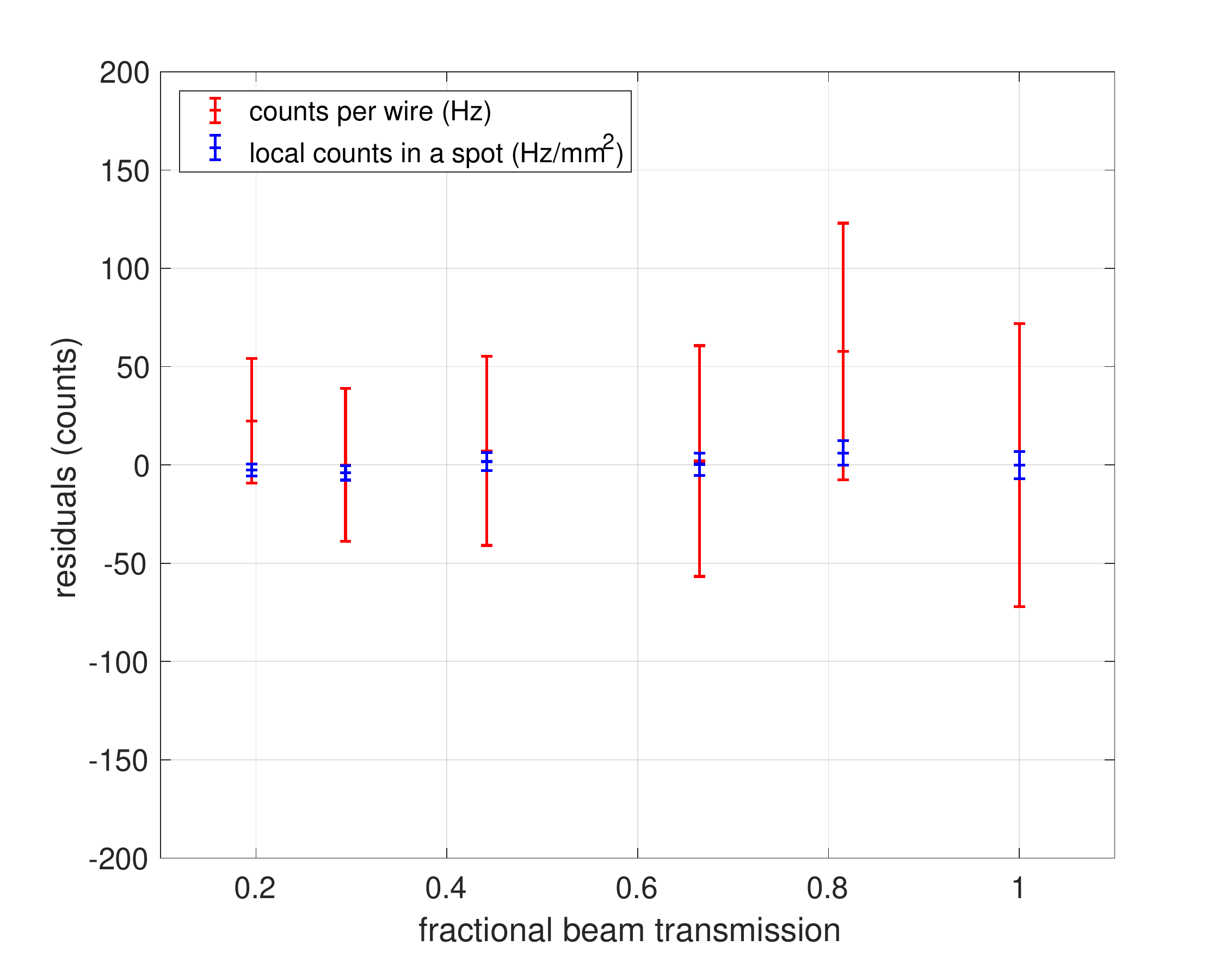}
\caption{\label{fig74} \footnotesize The count rate plotted as a function of the increasing beam flux on the detector (left) for a spot illumination (blue) and slit illumination on a single wire, i.e. channel (red) and the residuals from the fit (right).}
\end{figure}

\begin{figure}[htbp]
\centering
\includegraphics[width=.49\textwidth,keepaspectratio]{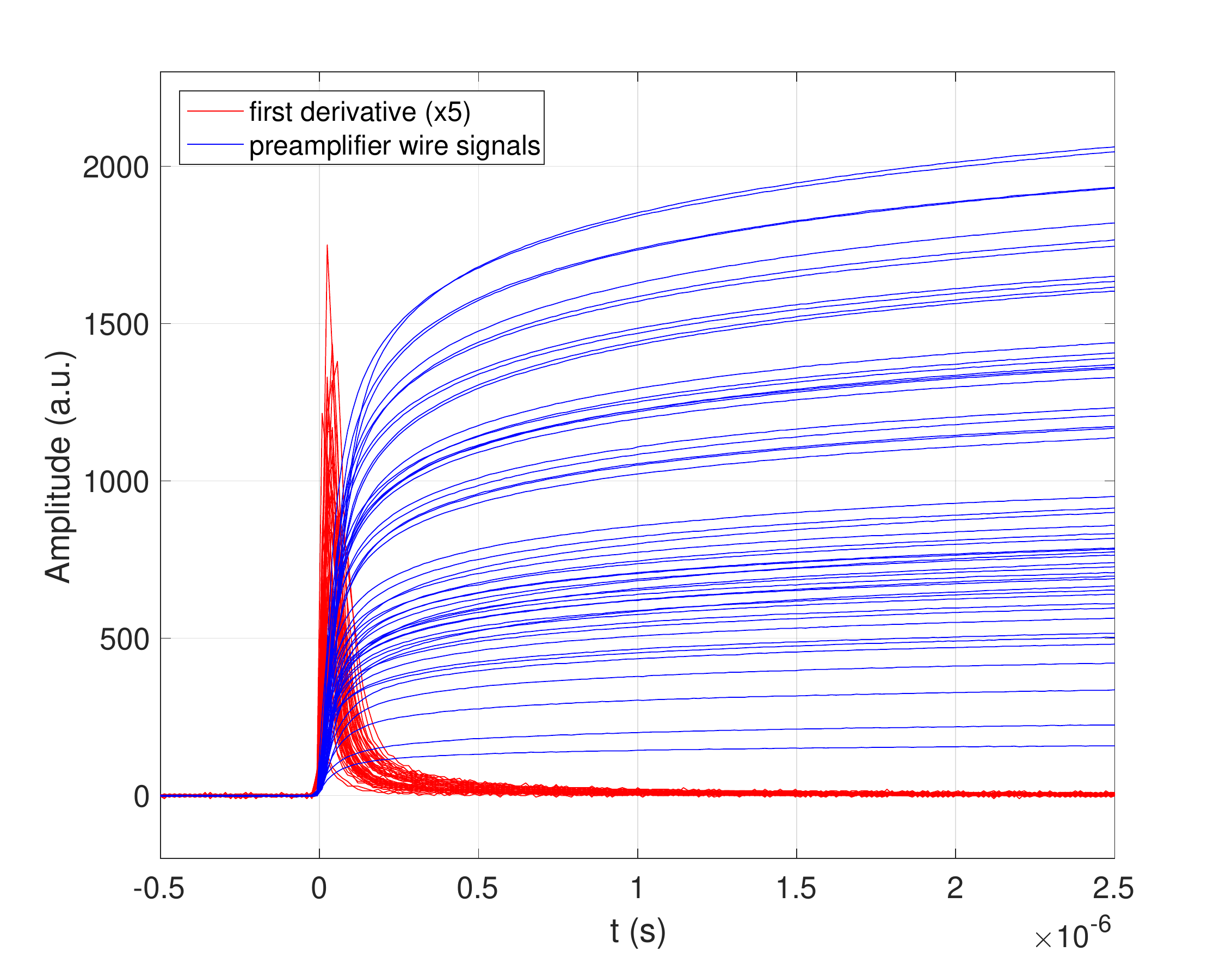}
\includegraphics[width=.49\textwidth,keepaspectratio]{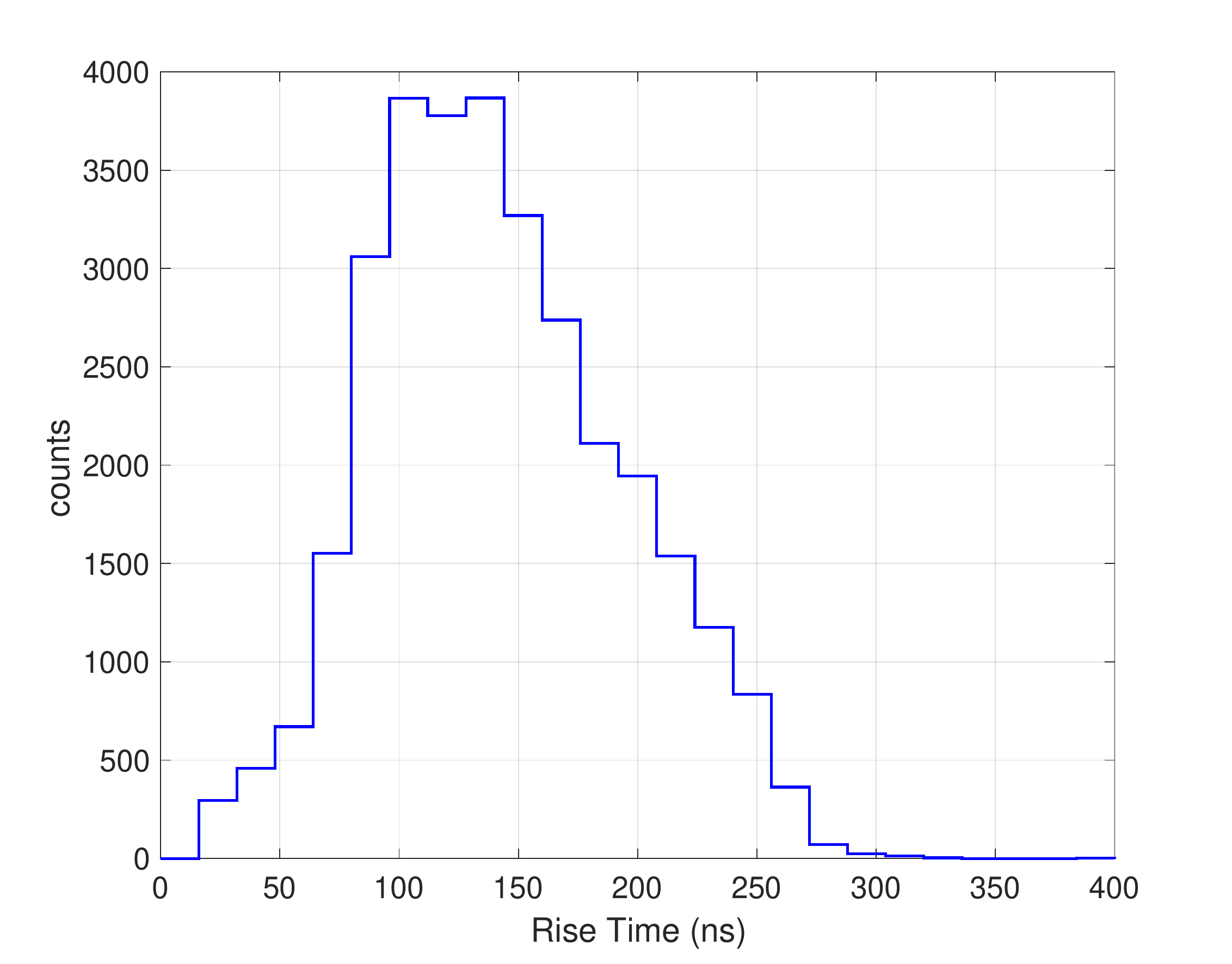}
\caption{\label{fig75} \footnotesize The pre-amplifier signals (without shaper) in blue and their first derivative (x5) in red (left) for a wire belonging to the cassette read out with individual amplifiers. The distribution of the rise time of the signals, i.e. their fast component (right).}
\end{figure}

For the measurement of the counting rate capability of the detector, the cassette equipped with individual amplifiers was used. Since we want to quantify the sole capability of the Multi-Blade without being limited by the DAQ, the front-end electronics is directly coupled to a constant fraction discriminator and a scaler (see Figure~\ref{figex1}). 
\\ The Multi-Blade was placed in the focal point of the monochromator in order to get the highest flux possible and the neutron beam was collimated on the cassette of interest by using two collimation slits. Many layers of a scattering material (A4 printer sheets without ink) were placed before the first collimation slit and they were progressively removed increase the neutron flux on the detector, up to the full transmission of the beam. Figure~\ref{fig74} shows the count rate for a single channel of the Multi-Blade as a function of an increasing beam intensity; a beam transmission of $1$ corresponds to a full transmission of the beam without any scatterer. Each point is obtained with a different amount of scattering material. The measurement is performed for a collimation in a spot to probe the local counting rate capability and in a vertical slit to probe the rate capability for a channel. The plot also shows the residuals of the measured point and the linear trend. 
A deviation from the linearity in the trend would suggest a space charge accumulation. As expected, we can clearly see that there is no sign of saturation up to the maximum measured rates which are $\left(1586\pm7\right)Hz/mm^2$ locally and $\left(16590\pm70\right)Hz$ per channel. Further tests are needed to prove the limit of the Multi-Blade and a more intense source must be involved. 
\\ The cassette with individual readout is then connected to the digitiser and the shaper board is omitted (see Figure~\ref{figex1}). We thus access the pre-amplifier signals directly in order to quantify the minimum shaping time needed for the signals. An example of the signals is shown in Figure~\ref{fig75} (left) and their first derivative is also shown to highlight their fast component. The distribution of the fast component of the signals, i.e. the rise time, is shown in the plot on the right in Figure~\ref{fig75}. The fast rise of the signals is due to the motion of the ions generated in the avalanche process that are travelling toward the cathodes. Although the ion collection time can be of the order of a few hundred microseconds, the pulse formation is generally $10^{-3}$ faster~\cite{DET_ravazzani}. The ions generate the most of the signal while they are moving in the higher intensity electric field region; i.e. close to the wire. For the Multi-Blade this fast component is at most $300\,ns$. 
\\With a shaping time of $500\,ns$ a post shaping signal of about $1.5\,\mu s$ can be obtained. For pile-up rejection purposes it corresponds to rates up to about $600\,KHz$ per channel.
\subsection{Images}
\begin{figure}[htbp]
\centering
\includegraphics[width=.7\textwidth,keepaspectratio]{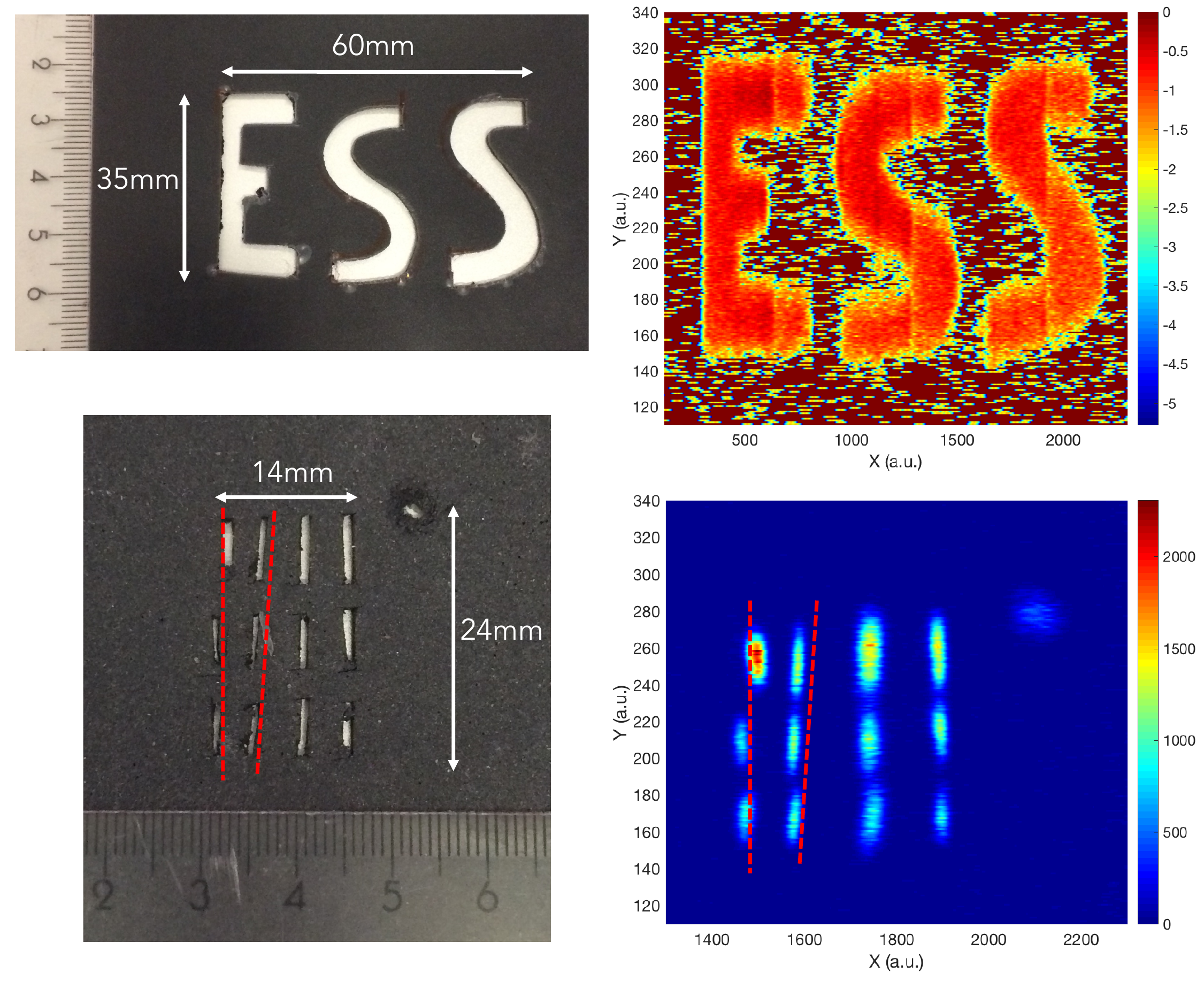}
\caption{\label{fig76} \footnotesize The boron masks used to obtain the images with the Multi-Blade (left) and the reconstructed images (right). The ESS-image is shown in logarithmic scale and the slit-image in linear scale.}
\end{figure}

A spread beam diffused by using a polyethylene scatterer was centred on the detector entrance window and a mask shown in Figure~\ref{fig76} was placed on it to obtain an image. The reconstruction was obtained with the 8 cassettes read out with charge division. Figure~\ref{fig76} also shows the reconstructed histograms, i.e. images, for the two masks used. 
\subsection{Gamma sensitivity}
\begin{figure}[htbp]
\centering
\includegraphics[width=.49\textwidth,keepaspectratio]{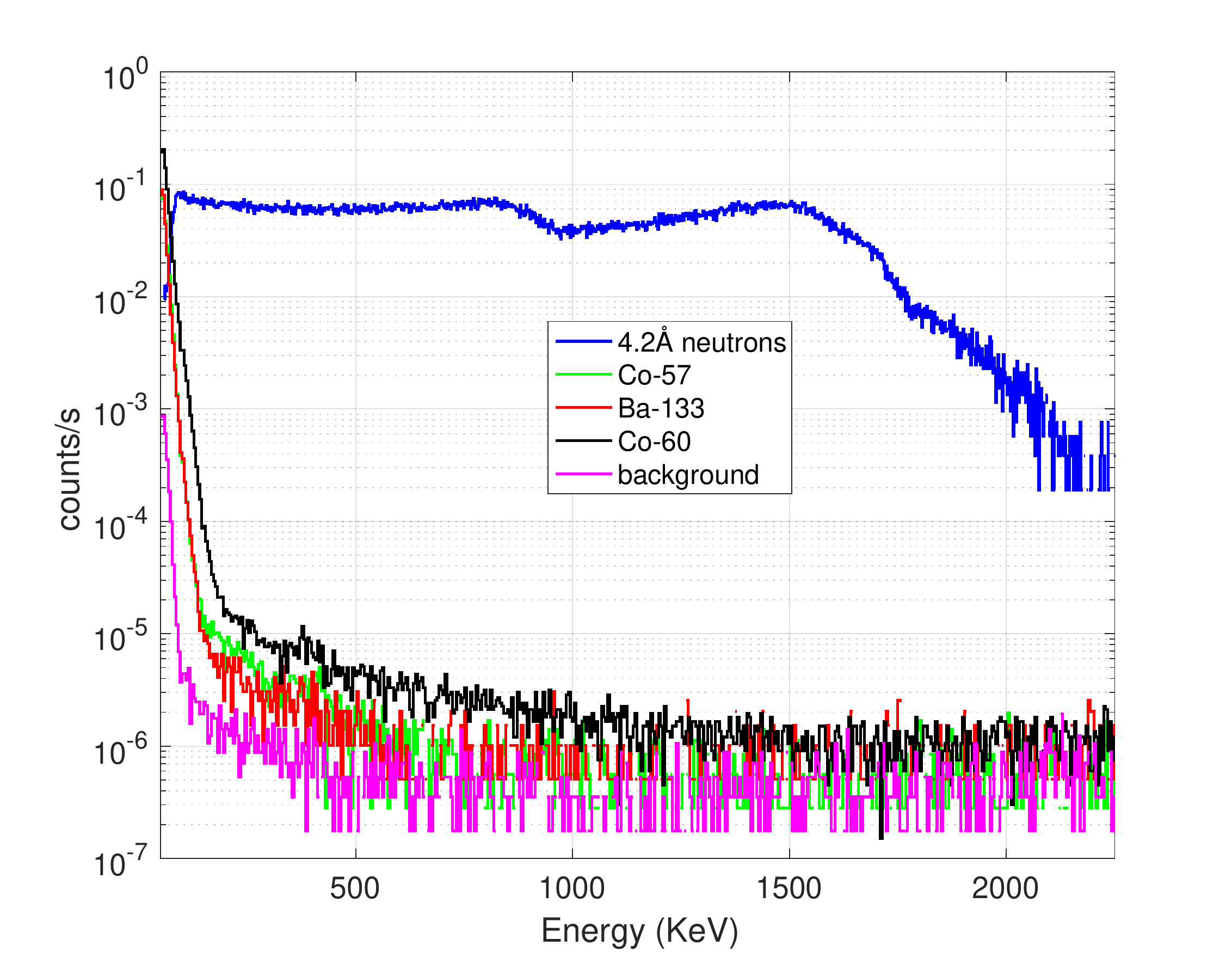}
\includegraphics[width=.49\textwidth,keepaspectratio]{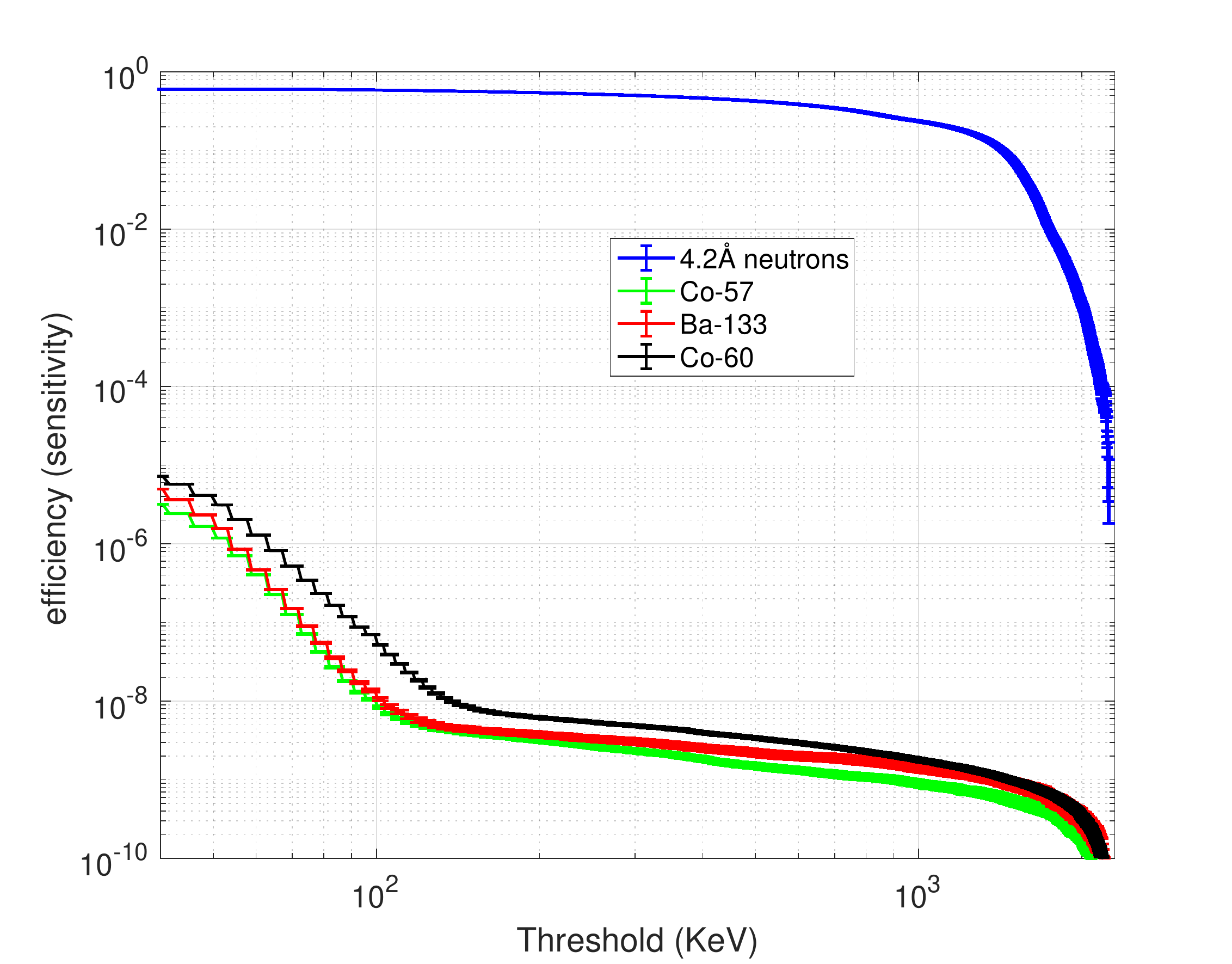}
\caption{\label{fig90} \footnotesize The PHS, normalized in time, recorded for the three $\gamma$-ray sources, for the background and for $4.2$\AA\, neutrons (left). The alpha peak of the $94\%$ branching ratio of the neutron capture reaction (corresponding to $1470\,KeV$) is used to convert the PHS X-axis from ADC levels to energy. The number counts in the PHS (normalized to the subtended solid angle and the activity of the sources) as a function of the applied threshold (right).}
\end{figure}

The gamma sensitivity of a neutron detector is defined as the probability for a $\gamma$-ray to generate a false count in a neutron measurement. The $\gamma$-ray sensitivity defines the best achievable signal-to-noise ratio, as in neutron scattering experiments the detectors are typically exposed to a large amount of $\gamma$-rays. It has been demonstrated that detector based on thin converter films, such as $\mathrm{^{10}B_4C}$ layers, have equal $\gamma$-ray rejection to that of $\mathrm{^3He}$ tubes. An extended study of the $\gamma$-ray sensitivity of neutron detectors based on thin converter films can be found in~\cite{MG_gamma}. 
To summarize, low energy photons (a few tens of $KeV$) generally transfer all their energy to an electron in a photoelectric interaction and they mainly interact with the gas of the detector, whereas medium energy photons (a few hundreds of $KeV$) Compton scatter losing only part of their energy and they mainly interact with the solid, i.e. the Aluminium as the mechanics. The energy for which the two contributions are of equal order of magnitude is $150\,KeV$ for $\mathrm{Ar/CO_2}$ at $1\,bar$ and Aluminium. Interaction with the gas are guaranteed to generate a signal while those in the solid only result in a signal if the electron reaches the gas. The deposited energy in the counting gas rarely exceeds a few tens of $KeV$ regardless of the energy of the photon; an electron of higher energy has a lower energy loss per unit track length. Only for several $MeV$ the measured energy increases. A low gas pressure is preferable for neutron detectors; to reduce the window thickness and to decrease the number of $\gamma$-ray converted. 
\\ For a gaseous detector the range of ions in the counting gas generated by the neutron capture is typically of the order of a few $mm$, while the range of electrons is much larger for the same energy. In detectors based on $\mathrm{^{10}B_4C}$ layers the ions from the neutron capture reaction can reach the gas after losing an arbitrary amount of energy while still in the layer; thus an arbitrarily low amount of energy can be deposited in the gas volume, therefore mixing with the photon spectrum. The neutron events that deposit an energy that can also be deposited by a $\gamma$-ray must be rejected. 
\begin{table}[htbp]
\centering
\caption{\label{tabgammasour} \footnotesize List of $\gamma$-ray sources used in the measurements with their activity (at the time of the measurement), main photons emitted and their relative intensity.}
\smallskip
\begin{tabular}{|l|c|c|}
\hline \hline
source & photon energy ($KeV$) & intensity ($\%$)\\
\hline
$^{57}Co$       & $6$    & $50$  \\
$(A = 7.01\cdot 10^6 \,Bq)$       & $7$ & $6$    \\
               & $14$    & $9.2$  \\
              & $122$    & $85.6$  \\
               & $136$    & $10.7$  \\
\hline
$^{133}Ba$       & $4$    & $15.7$  \\
$(A = 3.11\cdot 10^6 \,Bq)$       & $30.6$ & $33.9$    \\
               & $30.9$    & $62.2$  \\
              & $34.9$    & $17$  \\
               & $36$    & $3$  \\
              & $53$    & $2$  \\
               & $79$    & $2.6$  \\
             & $81$    & $32.9$  \\
               & $276$    & $7$  \\
               & $303$    & $18$  \\
              & $356$    & $62$  \\
              & $383$    & $9$  \\
\hline
$^{60}Co$       & $1173$    & $99.85$  \\
$(A = 1.33\cdot 10^7 \,Bq)$       & $1332$ & $99.98$    \\
\hline \hline
\end{tabular}
\end{table}
\\ The Multi-Blade response to reference $\gamma$-ray sources with known activity has been measured. The thresholds (hardware threshold) were set so that the electronic noise was rejected. Three sources were used and they are listed in Table~\ref{tabgammasour} with the main $\gamma$-rays emitted and their intensities. The activity at the time of the measurement is also shown. The $^{57}Co$ and $^{133}Ba$ can be considered as low energy sources whereas the $^{60}Co$ emits only photons above $1\,MeV$. 
\\ The $\gamma$-ray efficiency (or sensitivity) is defined as the probability for a photon incident on a detector element (a cassette or a single wire/strip) to result in an event. Practically the number of events that exceed a set threshold is normalized to the activity of the source and the solid angle. We define the flux incident on a detector element as all the photons emitted into the solid angle of that detector element. 
\\ Each source was placed close to the entrance window of the Multi-Blade, directly outside of the cassette equipped with the individual readout electronics; it results in a solid angle acceptance of approximately $0.2\,sr$. 
\\ A PHS for the three sources was recorded, and in absence of any source to estimate the background. The latter is the sum of the environmental background and the false counts that arises from the natural decay of radioactive contaminants typically present in standard Aluminium alloys~\cite{MG_bg}. In order to the get the PHS we sum all the counts recorded with the wires from $10$ to $30$ of the cassette in order to increase the statistics and to take into account only wires with approximately the same gas gain. Note that the wires from 1 to 9 have been excluded because of the strong gain variation (Section~\ref{simu}). The Multi-Blade is operated at $800\,V$. For each event the sum of all the wires involved in the detection process is shown and the event is associated to the wire with the largest amplitude, as for the neutron PHS in red in Figure~\ref{fig71}; the latter is also reported in Figure~\ref{fig90} (blue curve) for comparison. The alpha peak of the $94\%$ branching ratio of the neutron capture reaction (corresponding to $1470\,KeV$) is used to convert the PHS X-axis from ADC levels to energy. Figure~\ref{fig90} (left) shows the PHS recorded for each source and normalized in time, but not in solid angle and activity. Hence the $^{60}Co$ PHS shows a larger amount of counts because of the larger activity. 
\\ The average background rate in a single wire of the cassette used is approximately $7.8\cdot10^{-3}\,Hz$ if only the hardware threshold is applied just above the electronic noise, if a threshold of $100\,KeV$ is applied the total background rate in the PHS are $2.6\cdot10^{-4}\,Hz$. If a source is placed in front of the detector the rate in the PHS are $0.65\,Hz$, $0.75\,Hz$ and $3.38\,Hz$ for $^{57}Co$, $^{133}Ba$ and $^{60}Co$ respectively. If a $100\,KeV$ energy  is applied they decrease to $1.2\cdot10^{-3}\,Hz$, $1.3\cdot10^{-3}\,Hz$ and $1.9\cdot10^{-3}\,Hz$ respectively. 
\\ Figure~\ref{fig90} (right) shows the total counts in the PHS which is normalized to the activity of the source and the solid angle as a function of the threshold. Note that the statical errors are reported in the plot ,but the greatest uncertainty in this measurement is the measurement of the distance from the source to the sensitive element. We estimate that this can lead to the uncertainty of no more than a factor $2$. The encapsulation of the reference sources is the same and they were placed in the same position in front of the detector. Any error due to the distance uncertainty is the same for each point of the measurement, thus it does not impact the shape of the curves but instead results into an overall normalization shift. 
\\ It has been found that for any of the sources used a threshold of $100\,KeV$ lead to a $\gamma$-ray sensitivity (efficiency) below $10^{-7}$ per wire; whereas the neutron efficiency (at $4.2$\AA) for the same threshold is $\left( 56 \pm 2\right)\%$  (Section~\ref{effsusect}). Intuitively, since the threshold are set individually on each wire (or strip), any electron that travels a long distance in the cassette will spread its energy among many wires (or strip); i.e. the energy collected per wire (or strip) is a small quantity of the full deposited energy. 
\section{Conclusions and outlook}
The Table~\ref{tabref} summarizes the results of the test performed with the improved Multi-Blade detector (Multi-Blade 2016) and the results that had been obtained with the first demonstrator~\cite{MIO_MB2014,MIO_MBproc,MIO_MyThesis} (Multi-Blade 2013). 
\begin{table}[htbp]
\centering
\caption{\label{tabref} \footnotesize Multi-Blade detector performance from~\cite{MIO_MB2014,MIO_MBproc,MIO_MyThesis} and the new detector.}
\smallskip
\begin{tabular}{|l|l|l|}
\hline
\hline
 & Multi-Blade 2013 & Multi-Blade 2016 \\
\hline
\hline
gas gain & 58 & $20 \pm 3$ \\ 
\hline
efficiency & $\left(26 \pm 0.2 \right)\%$ \quad at ($10^{\circ}$, 2.5\AA) &  ${\it 44\%}$ \quad at ($5^{\circ}$, 2.5\AA) (calculated) \\
& & $\left(56 \pm 2 \right)\%$ \quad at ($5^{\circ}$, 4.2\AA) \\
& & $\left(65 \pm 2 \right)\%$ \quad at ($5^{\circ}$, 5.1\AA) \\
\hline
 spatial resolution  &  $\approx 0.3\times4\,mm^2$  & $\approx 0.5\times2.5\,mm^2$ \\ 
\hline
uniformity   &  $\leq 2\%$   &  $\leq 10\%$  \\ 
\hline
 overlap  &  $\approx 50\%$ eff. drop in $2\,mm$ gap  & $\approx 50\%$ eff. drop in $0.5\,mm$ gap \\ 
\hline
stability   &  --  & $ < 1\%$ in 12h \\ 
\hline
counting rate capability  &  --  & $>1.6\,KHz/mm^2$ \\ 
 & -- & $>16.6\,KHz/channel$ \\
 \hline
 gamma sensitivity & -- & $<10^{-7}$ \\
   & -- & (for $100\,KeV$ threshold) \\
\hline
\hline
\end{tabular}
\end{table}
The Multi-Blade is a promising alternative to $\mathrm{^3He}$ detectors for neutron reflectometry instruments at ESS and at other neutron scattering facilities in the world. The design of the Multi-Blade has been improved with respect to the past~\cite{MIO_MB2014} to meet the requirements for the ESS reflectometers. A demonstrator has been assembled at ESS and tested at the Budapest reactor (BNC~\cite{FAC_BNC}) and at the Source Testing Facility (STF)~\cite{SF1} at the Lund University (Sweden).
\\ The amount of material that can cause neutron scattering, and thus misaddressed events in the detector, has been completely removed when comparing to the previous prototype~\cite{MIO_MB2014}. The $\mathrm{^{10}B_4C}$-layer in the current detector serves as a shielding for the readout system, thus it can not be reached at all by neutrons. 
\\ The detector is operated at atmospheric pressure with a continuous gas flow. Cost-effective materials can be employed because out-gassing is not an issue. Moreover, because of the lower pressure with respect to $\mathrm{^3He}$ high pressure vessel, a thinner detector entrance window can be used resulting in a lower scattering. The detector was stable for $12h$ within $1\%$ variation. Longer stability tests are needed in the future. 
\\ The efficiency has been measured at an inclination of $5$ degrees to $\approx 56\%$ at 4.2\AA. The measured efficiency meets the requirements as well as the theoretical calculations; it is $\approx 44\%$ at 2.5\AA, the shortest neutron wavelength used in reflectometry at ESS.
\\ The high spatial resolution below $0.5\,mm$ has already been shown in the past~\cite{MIO_MB2014}. We found $0.5\,mm \times 2.5\,mm$ with the new detector. The spatial resolution depends on the algorithm that is used to reconstruct the event position and, due to the different geometry of the electric field of strips with respect to wires, the algorithm used affects to a larger extent the spatial resolution given by the cathode strips than that of the anode wires. The spatial resolution of the Multi-Blade is beyond of what can be achieved with the the state-of-the-art technologies in currents facilities at present. 
\\ Previously, a relative efficiency drop of $\approx 50\%$ was found in the previous detector within a gap of $2\,mm$ between the cassettes~\cite{MIO_MB2014}. However, the here presented improved detector geometry with a gap of $0.5\,mm$ resulted in approximately the same relative drop in efficiency. 
\\ It has been shown that this detector can be efficiently operated at a very low gas gain ($\approx 20$), a feature that is needed to reduce the space charge effects and improve the counting rate capability. 
\\ A uniformity within $20\%$ was found if the detector is operated at higher gain, which is needed for the charge division readout, it improves to $10\%$, as the electric field decreases with the individual readout which requires a lower gas gain. The current detector employs $2\,mm$-thick Al-substrates on which the $\mathrm{^{10}B_4C}$-layer is deposited, it has been already observed a deviation form planarity due to the residual stress after the deposition process. These Al-blades are replaced with Ti-blades in the next prototype. We have observed that the Ti-substrates do not show any visible curvature after deposition and therefore improve the detector uniformity.
\\ At the price of a more complex electric field geometry the strip cathode is shielded by the thick coating and thus prevented from causing scattering and misaddressed events in the detector. Measurements and simulations show a drop in gas gain at the front of each cassette; i.e. at the first 7 frontal wires. There are mainly three ways to correct this drop in gas gain  and they will be investigated in the next Multi-Blade generation. The wire thicknesses for the first 7 wires can be changed according to the gain drop in order to compensate the reduction, otherwise the gain drop can be compensated with a separate high voltage supply or by adjusting individual threshold on each channel. 
\\ The counting rate capability of the detector has been measured up to the highest rate available at the beam line. No saturation has been observed up to $\left(1586\pm7\right)Hz/mm^2$ locally and $\left(16590\pm70\right)Hz$ per single channel. 
\\ The $\gamma$-ray sensitivity of the Multi-Blade has been measured for three $\gamma$-ray sources ($^{57}Co$, $^{133}Ba$ and $^{60}Co$) and it was found that for an energy threshold of $100\,KeV$, which leads to the full neutron efficiency, it is below $10^{-7}$ for any photon between a few $KeV$ to approximately a $MeV$. 
\\ We are assembling a Multi-Blade detector which employs Ti-blades to get to the required electric field uniformity and it will be equipped only with individual readout over 576 channels. We intend to perform the standard detector characterization measurements and to take this detector to an existing reflectometer to measure some reflectometry standard samples in a real environment and compare the results with the state-of-the art technology.

\acknowledgments This work is being supported by the BrightnESS project, Work Package (WP) 4.2 (Horizon 2020, INFRADEV-3-2015, 676548) and carried out as a part of the collaboration between the European Spallation Source (ESS - Sweden), the Lund University (LU - Sweden), the Link\"{o}ping University (LiU - Sweden) and the Wigner Research Centre for Physics (Hungary).
\\ The work was supported by the Momentum Programme of the Hungarian Academy of Sciences under grant no. LP2013-60.
\\ The work was carried out in part at the Source Testing Facility, Lund University (LU - Sweden).
\\ The work originally started in the context of the collaboration between the Institut Laue-Langevin (ILL - France), the Link\"{o}ping University (LiU - Sweden) and the European Spallation Source (ESS - Sweden) within the context of the International Collaboration on the development of Neutron Detectors (www.icnd.org).

\bibliographystyle{ieeetr}
\bibliography{BIBLIODB}
\end{document}